\documentclass[twocolumn,aps,prd,superscriptaddress,showpacs,floatfix,longbibliography,10pt]{revtex4-1}
\usepackage{url}
\usepackage{cancel}
\usepackage[colorlinks,linkcolor=blue,citecolor=blue,filecolor=black,urlcolor=blue]{hyperref}
\usepackage{booktabs} 
\usepackage{tabularx} 
\usepackage{epsfig,graphics}
\usepackage{graphicx}
\usepackage{dcolumn}
\usepackage{bm}
\usepackage[usenames]{color}
\usepackage{amssymb}
\usepackage{amsmath}
\usepackage{slashed}
\usepackage{multirow}
\usepackage{float}
\usepackage{harpoon}
\usepackage{MnSymbol}
\usepackage{appendix}
\usepackage{color}
\usepackage{xcolor}
\usepackage{hyperref}
\usepackage{cleveref}
\usepackage{ulem} 
\usepackage{subcaption} 
\usepackage[justification=raggedright]{caption} 

\begin{document}


\title{%
  \texorpdfstring{The $\theta$-term effects on isospin asymmetric hot and dense quark matter}
  {The theta-term effects on isospin asymmetric hot and dense quark matter}
}
\author{Lei Zhang}\email[Correspond to\ ]{zhanglei231@mails.ucas.ac.cn}
\affiliation{School of Physical Sciences, University of Chinese Academy of Sciences, Beijing 100049, China}
\author{Lu-Meng Liu}\email[Correspond to ]{liulumeng@fudan.edu.cn}
\affiliation{Physics Department and Center for Particle Physics and Field Theory, Fudan University, Shanghai 200438, China}
\author{Mei Huang}\email[Correspond to\ ]{huangmei@ucas.ac.cn}
\affiliation{School of Nuclear Science and Technology, University of Chinese Academy of Sciences, Beijing, 100049, P.R. China}
\date{\today}

\begin{abstract}
We investigate the impact of the CP-violating $\theta$ term on isospin symmetry breaking in quark matter and compact star properties using a two-flavor Nambu-Jona-Lasinio (NJL) model. By incorporating the $\theta$ parameter through the Kobayashi-Maskawa-'t Hooft (KMT) determinant interaction, we derive the thermodynamic potential and gap equations under finite temperature, baryon chemical potential, and isospin chemical potential. At zero temperature and baryon density, $\theta$ suppresses conventional chiral ($\sigma$) and pion ($\pi$) condensates while promoting pseudo-scalar ($\eta$) and scalar-isovector ($\delta$) condensates, thereby reducing the critical isospin chemical potential $\mu_I^{\text{crit}}$ for spontaneous symmetry breaking. For $\theta=\pi$, a first-order phase transition emerges at $\mu_I^{\text{crit}} = 0.021$ GeV, accompanied by CP symmetry restoration. Extending the investigation to finite temperature and baryon chemical potential reveals that these $\theta$-term-induced effects persist. Axion effects (modeled via $\theta\equiv a/f_a$) stiffen the equation of state (EOS) of non-strange quark stars, increasing their maximum mass and radii, in agreement with multimessenger constraints from pulsar observations and gravitational wave events. These results establish $\theta$ as a critical parameter modulating both the Quantum Chromodynamics (QCD) phase structure and compact star observables.
\end{abstract}
\maketitle
\section{Introduction}
\label{sec:intro}
It is known that the strong interaction largely obeys the space-time reflection symmetry (P and T symmetry). However, this is not a direct consequence of Quantum Chromodynamics (QCD). As is well known, the instanton configurations existing in gauge fields and their close connection with the axial anomaly permit the existence of the CP-violating $\mathcal{L}_\theta$ term in the QCD Lagrangian:
\begin{equation}
    \mathcal{L}_\theta=\theta\frac{g^2}{32\pi^2}G\tilde{G},
    \label{term}
\end{equation}
where $G$ and $\tilde{G}$ denote the gluonic field strength tensor and its dual. The parameter $\theta$ is a dimensionless real number that can take arbitrary values. This Chern-Simons term does not affect the classical equations of motion. Being consistent with Lorentz invariance and gauge invariance, it violates charge conjugation and parity unless $\theta=0$ mod $\pi$. Experimental constraints from the neutron electric dipole moment require $\theta$ to be extremely small \cite{Baker2006,Griffith2009,Parker2015,Graner2016,Yamanaka2017}, suggesting a possible explanation through the spontaneous breaking of a new symmetry—the Peccei-Quinn (PQ) symmetry \cite{PecceiQuinn1977PRL}. By endowing $\theta$ with a dynamical character and elevating it to the axion field, $\theta(x)\equiv a(x)/f_a$, where $f_a$ is the axion decay constant, PQ dynamics introduces an effective potential $V_{\text{eff}}(\langle \theta \rangle) \sim -\Lambda_{\text{QCD}}^4 \cos(\langle \theta \rangle)$. This potential drives the expectation value of the axion field toward zero in vacuum. Spontaneous PQ symmetry breaking then generates a pseudo-Goldstone boson—the axion \cite{diCortona2016,KimCarosi2010}. 
In the original axion models established by Peccei and Quinn \cite{PecceiQuinn1977PRL,PecceiQuinn1977PRD}, Weinberg \cite{Weinberg1978}, and Wilczek \cite{Wilczek1978}, the PQ symmetry breaking occurs simultaneously with electroweak symmetry breaking.  However, this scenario conflicts with observations of $K$ and $J/\psi$ meson decays \cite{Kim1987}. This conflict can be avoided if the PQ spontaneous symmetry breaking occurs at a higher energy scale, producing very light and weakly interacting axions.

The spontaneous PQ symmetry breaking predicts the smallness of $\theta$, providing an elegant mechanism to solve the strong CP problem. However, the microscopic origin of the vanishing or extreme smallness of the CP-violating term is not fully understood. At zero temperature and density, according to the Vafa-Witten theorem \cite{PhysRevLett.53.535}, spontaneous parity violation does not arise at $\theta=0$. Meanwhile, when $\theta=\pi$, even though CP is conserved, spontaneous CP violation can occur through the so-called Dashen phenomenon with the appearance of two degenerate CP-violating vacua \cite{Dashen1971,Witten1980}. It is well known that the existence of a non-zero $\theta$ value in the non-perturbative regime of QCD leads to a rich vacuum structure, causing condensates in the pseudoscalar channel and generating a more complex phase diagram for the strong interaction. Therefore, studying QCD systems with non-zero $\theta$ is particularly interesting. However, before conducting research on $\theta$-dependent strong interactions,  we must rigorously identify the specific physical scenarios where a non-zero $\theta$ value might exist, to avoid violating the fundamental assumptions of the PQ mechanism.

First, even if CP is not violated in the QCD vacuum, CP violation may occur in QCD matter at finite temperature or density. It has been suggested that hot matter produced in heavy-ion collisions could generate locally CP-violating metastable regions through sphaleron transition processes \cite{Kharzeev2010}. Such states with effective non-zero $\theta$ can decay via CP-odd processes \cite{KharzeevPisarskiTytgat1998}. In addition to high temperatures, non-central nuclear collisions can produce strong magnetic fields. A nonzero $\theta$ would cause a deviation between the left- and right-handed quarks, thereby generating an electromagnetic current along the magnetic field. This mechanism, known as the chiral magnetic effect (CME) \cite{Kharzeev2006}, would induce the charge separation results observed in STAR experiments \cite{Abelev2009}. However, for central collisions, recent studies indicate that local parity violation leads to the formation of pseudoscalar condensates, which may affect excess dilepton production in such collisions \cite{ANDRIANOV2012230}. In fact, Refs. \cite{MetlitskiZhitnitsky2005,KharzeevZhitnitsky2007} propose that $\theta$ could be of order one during the QCD phase transition in the early universe because sphalerons are sufficiently active to overcome the potential barrier between different degenerate vacuum states \cite{McLerranMottolaShaposhnikov1991}, while it vanishes in the present epoch \cite{PecceiQuinn1977PRD,Kim1979,ShifmanVainsteinZakharov1980,Zhitnitsky1980,DineFischlerSrednicki1981}.

Furthermore, as a leading dark-matter candidate, axions may influence compact star properties through gravitational capture during stellar formation and neutron star (NS) mergers (particularly binary NS mergers). Consequently, compact stars could accumulate substantial amounts of dark matter \cite{PhysRevD.105.023001}, in particular, in the axion form \cite{Lopes2022}. The presence of axions inside compact stars enhances interactions with strongly interacting matter, substantially altering stellar structure and thermodynamic properties, which could manifest in observational signatures \cite{PanotopoulosLopes2017,NelsonReddyZhou2019,Ellis2018,IvanytskyiSagunLopes2020,Rezaei2017,deLavallazFairbairn2010,GreshamZurek2019}. The self-annihilation of accumulated dark matter in compact stars increases luminosity and temperature \cite{Kouvaris2008,FullerOtt2015,AcevedoBramanteGoodman2021}, and modifies the cooling curves of stars with specific masses \cite{Sedrakian2019,Sedrakian2016}.

Incorporating the axion into the models is analogous to introducing the $\theta$ angle in QCD. In fact, formally one can transition from QCD with a finite $\theta$ to QCD with a finite axion background. Therefore, studying the interaction between QCD and axions is equivalent to investigating finite $\theta$. In our work, we will use $\theta$ and $a/f_a$ interchangeably.

Due to the non-perturbative nature of QCD, studying the QCD phase structure with arbitrary $\theta$ values in the QCD Lagrangian is extremely challenging. In this regime, effective models work alongside first-principle lattice QCD calculations in exploring non-perturbative physics. Consequently, the effects of the theta term (simply called the $\theta$ effects or CP-violating effects) in the strong interaction have been extensively studied using low-energy effective theories such as chiral perturbation theory \cite{Smilga1999,Tytgat2000,AkemannLenaghanSplittorff2002,Creutz2004,MetlitskiZhitnitsky2005}, the linear sigma model \cite{MizherFraga2009a}, the Nambu-Jona-Lasinio (NJL) model and its various extensions \cite{FujiharaInagakiKimura2007,BoerBoomsma2008,BoerBoomsma2009,SakaiKounoSasakiYahiro2011,PhysRevD.100.076021,PhysRevD.110.014042,PhysRevD.100.014013}. Specifically, CP violation has been investigated in chiral phase transitions \cite{SakaiKounoSasakiYahiro2011,PhysRevD.103.074003,PhysRevD.85.114008}, chiral phase transitions under magnetic backgrounds \cite{PhysRevD.91.034031}, and effects in color superconducting phases \cite{PhysRevD.110.014042}. Our work focuses on using the NJL model to study $\theta$ effects at finite isospin density. The effects at a finite isospin density without considering CP violation have already been extensively studied \cite{PhysRevD.71.116001,Liu:2021gsi,Liu:2023uxm,XU2021115540,HE200593,Lu2020,Lu2021}. When spontaneous isospin symmetry breaking occurs with non-zero $\theta$,  both scalar and pseudo-scalar meson condensates ($\delta$, $\pi$) will be induced.

Modulation of the QCD phase structure by the $\theta$ term will affect the bulk properties of quark matter, thereby influencing the equation of state (EOS) of strongly interacting matter. The EOS largely determines the structure of neutron stars and quark stars. Given a specific EOS, the corresponding mass-radius relations can be obtained by solving the Tolman-Oppenheimer-Volkoff (TOV) equations. At very low temperatures, quarks remain confined in hadrons at low chemical potentials but become deconfined at high chemical potentials. Therefore, highly dense pulsars are more likely to be quark stars rather than neutron stars. Based on the hypothesis that strange quark matter might be the true ground state of strongly interacting matter \cite{Itoh1970,Bodmer1971,Witten1984}, many authors have conducted extensive studies on the properties of strange quark stars, including pure quark stars and hybrid neutron stars with quark cores \cite{FarhiJaffe1984,Li2017,Li2018,Li2019,Li2020,Sedaghat2022,Geng2021}. A recent study suggests that stable quark matter might not be strange when taking the flavor-dependent feedback of the quark gas on the QCD vacuum into account \cite{Holdom2018}, implying that non-strange quark stars could exist. Subsequent studies on non-strange quark stars have been based on this viewpoint \cite{Wang2019,Zhao2019,Ren2020,ZhangMann2021}. Furthermore, research shows that under combinations of vector interactions and exchange interactions in quark matter, both non-strange and strange quark matter could be stable \cite{Yuan2022}. Thus, whether two-flavor or three-flavor quark matter is more stable remains an open question. Similar to previous studies \cite{PhysRevD.111.L051501,Kumar:2024abb}, in this work we will thoroughly investigate the properties of non-strange quark stars containing axions to reflect the $\theta$ effects on the bulk properties of quark matter.

In the present work we focus our attention on how isospin transition is affected when there is a $\theta$ term in the Lagrangian. For this purpose, we adopt the two-flavor NJL model as an effective theory for isospin symmetry breaking in strong interaction. The CP violating parameter $\theta$ is included in the Kobayashi-Maskawa-t’ Hooft (KMT) determinant term. In this context we note that the two-flavor scenario for spontaneous CP violation for $T$-$\mu_B$ has been studied in this model \cite{SakaiKounoSasakiYahiro2011}. This will be further extended to study the restoration of CP at finite isospin chemical potential .

This paper is organized as follows. In Sec.~\ref{sec:theory}, we establish the mean field theory of the NJL model at finite temperature and baryon and isospin densities and $\theta$. In Sec.~\ref{sec:results1}, we study the $\theta$ effects on the isospin symmetry breaking at zero temperature, zero baryon chemical potential, but finite isospin chemical potential. In Sec.~\ref{sec:results2}, we study the temperature behavior of condensates in the isospin symmetry breaking phase at zero baryon chemical potential and show the phase diagrams in the temperature and isospin chemical potential plane with nonzero $\theta$. And we extend phase diagram for CP transition in the $T-\mu_I$ plane. In Sec.~\ref{sec:results3}, we discuss the $\theta$ effects at finite temperature and baryon density in the isospin symmetry breaking phase. In Sec.~\ref{sec:results4} we briefly discuss the effects of  QCD axion on the nonstrange quark star properties. We conclude and outlook in Sec.~\ref{sec:summary}.

\section{\texorpdfstring{%
  $\theta$ Within the NJL Model at Finite Isospin Density
}{%
  theta Within the NJL Model at Finite Isospin Density
}}
\label{sec:theory}
\subsection{The Lagrangian}
Given that we employ the NJL model to investigate the non-perturbative effects of the topological term, it is necessary to briefly review this theoretical framework, which describes the interactions between quarks,  and effectively incorporates the topological term of the gluon field in the expression. We now review the physical correspondence between instanton effects, the $U(1)_A$ anomaly, and the topological term in QCD with the Kobayashi-Maskawa-'t Hooft term in the NJL model to demonstrate the validity of the NJL framework \cite{PhysRevLett.37.8,Hooft1976,Hooft1986}.

As the fundamental theory of strong interactions, QCD exhibits a crucial feature—the $U(1)_A$ anomaly. At the classical level, the QCD Lagrangian remains invariant under chiral transformations. However, in the quantum theory, the fermionic integration measure in the path integral acquires a phase variation under chiral rotations ($\psi \rightarrow e^{i\alpha\gamma_5}\psi$). Through the Fujikawa's method, this measure variation is calculated as
\begin{equation}
\mathcal{D}\psi\mathcal{D}\bar\psi\rightarrow\mathcal{D}\psi\mathcal{D}\bar\psi \exp\left[-i\alpha\frac{N_fg^2}{16\pi^2}\int d^4xG\tilde{G}\right].
\end{equation}
This phase factor effectively introduces an additional term $\Delta\mathcal{L}=-\alpha\frac{N_fg^2}{16\pi^2}G\tilde{G}$ in the Lagrangian. To maintain quantum consistency, the original Lagrangian must explicitly contain the topological term Eq.~(\ref{term}). The dependence on the coupling constant $g$ renders this term negligible in perturbative QCD but significant in non-perturbative regimes. This explicitly demonstrates the intrinsic connection between the topological term and the $U(1)_A$ anomaly.

At the quantum level, the chiral $U(1)_A$ symmetry implied by the QCD Lagrangian remains unrealized. The corresponding conserved current $j_5^\mu=\bar{\psi}\gamma^\mu\gamma^5\psi$ becomes non-conserved, satisfying in the chiral limit
\begin{equation}
    \partial_\mu j_5^\mu=\frac{N_fg^2}{16\pi^2}G\tilde{G},
    \label{current}
\end{equation}
where the right-hand side denotes the topological charge density. Its spacetime integral defines the instanton number (Pontryagin number)
\begin{equation}
    Q=\frac{g^2}{32\pi^2}\int_M d^4x G\tilde{G}.
\end{equation}
As an integer-valued topological invariant, $Q$ originates from the mathematical structure of gauge fields. For SU($N$) gauge groups, $Q$ is classified by the third homotopy group $\pi_3(SU(N))=\mathbb{Z}$, representing the Chern number of the fiber bundle topology. This classification describes how gauge field configurations "wrap" around Euclidean spacetime. The non-zero fluctuations of $Q$ establish the fundamental connection between the $U(1)_A$ anomaly and instantons. We further propose that instanton effects through non-perturbative processes (when $g$ is large) break $U(1)_A$ symmetry and generate topological charge contributions.

This analysis reveals the intrinsic connections among instanton effects, $U(1)_A$ anomaly, and the topological term. The critical challenge lies in constructing $U(1)_A$-violating interaction terms with instanton effects in the NJL model to match real QCD behavior. To achieve this, we must first understand instanton-induced quark interactions.

In QCD, instantons modify quark chirality (left/right-handedness) and induce multi-flavor correlations through their topological structure. First, the chiral anomaly equation Eq.~(\ref{current}) causes net chirality change. Second, instanton gauge configurations support fermion zero modes—solutions to the Weyl equation in massless quark backgrounds with instanton number $Q=\pm 1$. The Atiyah-Singer index theorem guarantees zero-mode solutions for arbitrary $Q$, but higher $Q$ contributions are exponentially suppressed by the action $S_{\text{inst}} \propto |Q|$ in the path integral. Thus, $Q=\pm 1$ instantons dominate low-energy physics. The physical consequence of these zero modes is chirality flipping:  instantons with $Q=1$ flip left-handed quarks to right handed quarks, and $Q=-1$ anti-instantons flip right-handed quarks to left-handed quarks  through their chiral zero modes.   Crucially, this effect preserves flavor symmetry by simultaneously acting on all quark flavors. These considerations motivate the effective interaction term
\begin{equation}
    \mathcal{L}_{\text{eff}}\sim e^{-S_{\text{inst}}}\left(\prod_{f,f'}^{N_f}\bar\psi_{L,f}\psi_{R,f'}+\prod_{f,f'}^{N_f}\bar\psi_{R,f}\psi_{L,f'}\right),
\end{equation}
where $S_{\text{inst}}=\frac{8\pi^2}{g^2}$ denotes the single-instanton action. The factor $e^{-S_{\text{inst}}}$ weights instanton contributions, while $\prod_{f,f'}^{N_f}$ enforces flavor symmetry. The operator $\bar\psi_{L,f}\psi_{R,f'}$ represents the instanton-induced interactions which mediate the coupling between left-handed and right-handed fermions, leading to chirality flipping. Grassmann algebra properties necessitate full antisymmetrization of field operators for non-zero Grassmann integrals ($\mathcal{D}\psi^\dagger\mathcal{D}\psi$). Incorporating the topological term $\mathcal{L}_\theta$ in QCD introduces additional weighting in the effective potential. The topological term's action is
\begin{equation}
    S_{\theta}(Q)=\int_M d^4x \mathcal{L}_{\theta}=\theta Q.
\end{equation}
Thus, the complete effective Lagrangian becomes
\begin{align}
    \mathcal{L}_{\text{eff}}\sim &e^{-S_{\text{inst}}}\left(e^{-i S_\theta(+1)}\prod_{f,f'}^{N_f}\bar\psi_{L,f}\psi_{R,f'}+e^{-iS_\theta(-1)}\prod_{f,f'}^{N_f}\bar\psi_{R,f}\psi_{L,f'}\right) \notag \\
    =&e^{-S_{\text{inst}}}\left(e^{-i \theta}\prod_{f,f'}^{N_f}\bar\psi_{L,f}\psi_{R,f'}+e^{i\theta}\prod_{f,f'}^{N_f}\bar\psi_{R,f}\psi_{L,f'}\right).
\end{align}
Here, $\bar\psi_{L,f}\psi_{R,f'}$ corresponds to positive-instanton effects, while $\bar\psi_{R,f}\psi_{L,f'}$ to anti-instantons. Through determinant formation across quark flavors, We use determinant instead of $\prod_{f,f'}^{N_f}$ and recover the NJL model's renowned 't Hooft interaction term
\begin{equation}
    \mathcal{L}_{\text{det}}=-K\left[e^{-i\theta}\det\bar{\psi}(1+\gamma^5)\psi+e^{i\theta}\det\bar{\psi}(1-\gamma^5)\psi\right].
\end{equation}
This instanton-derived interaction simultaneously incorporates the topological term, breaks $U(1)_A$ symmetry, and preserves $SU(N_f)_L\otimes SU(N_f)_R$ symmetry. Remarkably, this mechanism dynamically generates the $\eta'$ mass within the NJL framework \cite{PhysRevLett.37.8}, predating the Witten-Veneziano formula in QCD \cite{WITTEN1979269,VENEZIANO1979213}.

The conventional method for introducing the $\theta$ term in 't Hooft interactions employs chiral rotations of the NJL Lagrangian, mirroring the QCD approach \cite{PhysRevD.100.014013}. This yields identical results to our path-integral derivation via topological term weighting.

We thus arrive at the two-flavor NJL Lagrangian for investigating CP violation
\begin{equation}
       \mathcal{L}_{\text{NJL}}=\bar{\psi}(i\slashed{\partial}-m)\psi+\mathcal{L}_{\bar{q}q}+\mathcal{L}_{\text{det}},
\end{equation}
where $\psi$ denotes the quark fields, $m$ the current quark mass, and the interaction terms are specified as
\begin{equation}
    \mathcal{L}_{\bar{q}q}=G\sum_{a=0}^3\left[ (\bar{\psi}\tau^a \psi)^2 +(\bar{\psi}i\gamma^5\tau^a \psi)^2 \right].
\end{equation}
Here, $\tau^a$ ($a=1,2,3$) are Pauli matrices and $\tau^0=I$, $G$ parametrizes four-quark scalar/pseudoscalar interactions, and $K$ controls the KMT determinant interaction. In some conventions, $K$ relates to an alternative coupling $G'$ through $K=-2G'$.

\subsection{The thermodynamic potential in the mean-field approximation}

To study the system at finite chemical potentials and temperature, we introduce the chemical potentials in the Lagrangian density
\begin{equation}
\mathcal{L}=\mathcal{L}_\text{NJL}+\bar{\psi}\hat{\mu}\gamma_0\psi,
\end{equation}
where $\hat{\mu}=\text{diag}(\mu_u,\mu_d)$ is the chemical potential matrix with
\begin{gather}
    \mu_u = \frac{\mu_B}{3} + \frac{\mu_I}{2}, \\
    \mu_d = \frac{\mu_B}{3} -\frac{\mu_I}{2}.
\end{gather}
where $\mu_u$, $\mu_d$ are the baryon, isospin potential, respectively.

Based on the mean-field approximation as detailed in Appendix~\ref{A}, the above Lagrangian density can be written as
\begin{equation}
    \mathcal{L}_\text{MF}=\bar{\psi}S^{-1}\psi-\nu,
\end{equation}
where
\begin{equation}
    S^{-1}(p)=\begin{pmatrix}
        \slashed{p}+\mu_u\gamma^0-M_u+\beta_{ud}i\gamma^5&S+Ri\gamma^5\\
        S+Ri\gamma^5& \slashed{p}+\mu_d\gamma^0-M_d+\beta_{du}i\gamma^5
    \end{pmatrix},
\end{equation}
is the inverse of the quark propagator $S^{-1}(p)$ as a function of quark momentum $p$, with
\begin{gather}
    \alpha_{ud}=\left(-4G\sigma_u+2K\cos(\frac{a}{f_a})\sigma_d-2K\sin(\frac{a}{f_a})\eta_d\right),\\
    \beta_{ud}=\left(4G\eta_u+2K\cos(\frac{a}{f_a})\eta_d+2K\sin(\frac{a}{f_a})\sigma_d\right),\\
    S=2G\delta-K\left[\sin(\frac{a}{f_a})\pi-\cos(\frac{a}{f_a})\delta\right],\\
    R=2G\pi-K\left[\cos(\frac{a}{f_a})\pi+\sin(\frac{a}{f_a})\delta\right],\\
    M_u=m_u+\alpha_{ud},\quad\quad M_d=m_d+\alpha_{du},
\end{gather}
and
     \begin{align}
  \nu =&\, G \bigg[ 2\left(\eta_u^2 + \eta_d^2\right) + 2\left(\sigma_u^2 + \sigma_d^2\right) + \pi^2 \bigg] \nonumber \\
  &- 2K \Bigg[ \cos\left(\frac{a}{f_a}\right) \bigg( \sigma_u\sigma_d - \eta_u\eta_d - \frac{\delta^2}{4} + \frac{\pi^2}{4} \bigg) \nonumber \\
  &\quad - \sin\left(\frac{a}{f_a}\right) \bigg( \sigma_d\eta_u + \sigma_u\eta_d - \frac{\delta\pi}{2} \bigg) \Bigg]
\end{align}
being the condensation energy independent of the quark fields. In most previous calculations, particularly those related to the QCD phase diagram, only the flavor-singlet scalar condensate $\sigma$ and non-singlet pseudoscalar condensate $\pi$ were considered. For the isospin-singlet pseudoscalar condensate $\eta$ (analogous to $\sigma$) and the non-singlet scalar condensate $\delta$ (analogous to $\pi$), these remain dynamically suppressed in the mean-field approximation without CP violation. This suppression arises because they enter the thermodynamic potential $\Omega$ through quadratic ($\eta^2$, $\delta^2$) and cross ($\eta\cdot\delta$) terms, which are not generated by the standard NJL interaction mechanism and align with real physical results.

In our study, the axion-axial current coupling introduces linear terms in $\Omega$ that explicitly source these condensates, necessitating their inclusion in the dynamical framework. This behavior qualitatively matches previous investigations of normal quark matter under CP-odd conditions \cite{PhysRevD.100.076021,Lopes2022,PhysRevD.100.014013,PhysRevD.103.074003}. Both scalar and pseudoscalar condensates must now be incorporated in the analysis. Below we show the explicit expressions on scalar and pseudoscalar condensates:
\begin{gather}
     \sigma_q= \langle \bar{\psi}_q \psi_q \rangle: \quad\sigma_u = \langle \bar{u} u \rangle,\quad\sigma_d = \langle \bar{d} d \rangle,\\
\pi^+ = \langle \bar{\psi} i \gamma_5\tau_+ \psi \rangle = 2 \langle \bar{u} i\gamma_5 d \rangle, \quad \pi^- = \langle \bar{\psi} i \gamma_5\tau_- \psi \rangle = 2 \langle \bar{d} i \gamma_5 u \rangle,\\
\eta_q=\langle \bar{\psi_q}i\gamma_5 \psi_q \rangle: \quad \eta_{u}=\langle \bar{u}i \gamma_5 u\rangle, \quad \eta_{d}=\langle \bar{d}i \gamma_5 d\rangle,\\
\delta^+ = \langle \bar{\psi} \tau_+ \psi \rangle = 2 \langle \bar{u}  d \rangle, \quad\quad \delta^- = \langle \bar{\psi}\tau_- \psi \rangle = 2 \langle \bar{d}  u \rangle.
 \end{gather}
The phase factor associated with the condensates $\pi^+$, $\pi^-$, $\delta^+$, and $\delta^-$ determines the spontaneous breaking direction of the $U(1)_I$ symmetry. However, this phase selection does not alter physical observables in globally thermalized systems \cite{PhysRevD.71.116001}. For computational simplicity, we fix the phase factor to zero in subsequent calculations, thus we take 
\begin{equation}
\pi^+=\pi^-=\pi,~~~  \delta^+=\delta^-=\delta.
\end{equation}  
Consequently, for any solution $(\pi, \delta)$ satisfying the gap equations, the configuration $(-\pi, -\delta)$ constitutes another valid solution to these equations.

Starting from the partition function as detailed in Appendix~\ref{B}, the thermodynamic potential can be expressed as
\begin{equation}
    \Omega=\nu-2N_c\sum_{i=1}^{4}\int\frac{d^3\vec{p}}{(2\pi)^3}\left[\frac{\lvert \tilde{E}_i\rvert}{2}+T\ln\left(1+e^{-\beta \lvert \tilde{E}_i\rvert}\right)\right],
    \label{thermo}
\end{equation}
where $\beta=1/T$ is the inverse of the temperature, and The effective quark energies in the above are defined $\tilde{E}_i=\lambda_i-\mu_B/3$ as detailed in Appendix~\ref{B}. In the present study without considering the color superconductivity, the color degree of freedom contributes a factor of $N_c =3$ and the factor 2 is comes out for the spin $1/2$ nature of the fermions.
\subsection{Gap equations}
Using the quark propagator as detailed in Appendix~\ref{C}, the expressions of the condensates in terms of the phase-space distribution function can be written as
\begin{gather}
    \sigma_q=4N_c \sum_{i=1}^4\int\frac{d^3\vec{p}}{(2\pi)^3}g_{\sigma q}(\tilde{E}_i)f(\tilde{E}_i),\label{gap1}\\
    \eta_q=4N_c \sum_{i=1}^4\int\frac{d^3\vec{p}}{(2\pi)^3}g_{\eta q}(\tilde{E}_i)f(\tilde{E}_i),\\
    \pi=4N_c \sum_{i=1}^4\int\frac{d^3\vec{p}}{(2\pi)^3}g_{\pi}(\tilde{E}_i) f(\tilde{E}_i),\\
    \delta=4N_c \sum_{i=1}^4\int\frac{d^3\vec{p}}{(2\pi)^3}g_{\delta}(\tilde{E}_i)f(\tilde{E}_i).
    \label{gap4}
\end{gather}
And the net-quark densities for $u, d$ quarks in terms of the phase-space distribution function can be written as
\begin{equation}
    \rho_q
=4N_c\sum_{i=1}^4\int\frac{d^3\vec{p}}{(2\pi)^3}g_{\rho q}(\tilde{E}_i)\left(-\frac{1}{2}+f(\tilde{E}_i)\right).
\label{density}
\end{equation}
with $q=u,d$ being the quark flavor. The net baryon density $\rho_B$ and the isospin density $\rho_I$ can be calculated from $\rho_B=\frac{1}{3}(\rho_u+\rho_d)$, and 
$\rho_I=\frac{1}{2}(\rho_u-\rho_d)$.
In the above expressions, $f(E)=\frac{1}{\exp(\beta E)+1}$
is the Fermi-Dirac distribution, the $g$ functions have the form of 
\begin{widetext}
\begin{gather}
    g_{\rho u}(\tilde{E}_i)=-\frac{\left[E_d^2-(\tilde{E}_i+\mu_d)^2+\beta_{d u}^2\right](\tilde{E}_i+\mu_d)+(S^2+R^2)(\tilde{E}_i+\mu_u)}{\prod_{k\neq i}(\tilde{E}_i-\tilde{E}_k)},
    \label{35}\\
    g_{\sigma u}(\tilde{E}_i)=\frac{-\left[E_d^2-(\tilde{E}_i+\mu_d)^2+\beta_{d u}^2\right]M_u+(S^2-R^2)M_d-2SR\beta_{du}}{\prod_{k\neq i}(\tilde{E}_i-\tilde{E}_k)},\\
    g_{\eta u}(\tilde{E}_i)=\frac{\left[E_d^2-(\tilde{E}_i+\mu_d)^2+\beta_{d u}^2\right]\beta_{ud}+2SR M_d+(S^2-R^2)\beta_{du}}{\prod_{k\neq i}(\tilde{E}_i-\tilde{E}_k)},\\
    g_{\rho d}(\tilde{E}_i)=g_{\rho u}(\tilde{E}_i)(u \leftrightarrow d),\quad g_{\sigma d}(\tilde{E}_i)=g_{\sigma u}(\tilde{E}_i)(u \leftrightarrow d),\quad g_{\rho d}(\tilde{E}_i)=g_{\rho u}(\tilde{E}_i)(u \leftrightarrow d),\\
    g_{\pi}(\tilde{E}_i)=2\frac{(S^2+R^2)R+S(M_u\beta_{du}+M_d\beta_{ud})+R\left[\vec{p}^2-(\tilde{E}_i+\mu_u)(\tilde{E}_i+\mu_d)+M_uM_d-\beta_{ud}\beta_{du}\right]}{\prod_{k\neq i}(\tilde{E}_i-\tilde{E}_k)},\\
     g_{\delta}(\tilde{E}_i)=2\frac{(S^2+R^2)S+R(M_u\beta_{du}+M_d\beta_{ud})+S\left[\vec{p}^2-(\tilde{E}_i+\mu_u)(\tilde{E}_i+\mu_d)-M_uM_d+\beta_{ud}\beta_{du}\right]}{\prod_{k\neq i}(\tilde{E}_i-\tilde{E}_k)}.\label{40}
\end{gather}
\end{widetext}
and they satisfy the following relations
\begin{equation}
    \sum_{i=1}^4g_\delta(\tilde{E}_i)= \sum_{i=1}^4g_{\pi}(\tilde{E}_i)= \sum_{i=1}^4g_{\sigma q}(\tilde{E}_i)= \sum_{i=1}^4g_{\eta q}(\tilde{E}_i)=0,\label{mathe}\\
\end{equation}
     \begin{equation}
          \sum_{i=1}^4g_{\rho q}(\tilde{E}_i)=1,\quad
     \sum_{q}g_{\rho q}(\tilde{E}_i)=\frac{1}{2}.
     \end{equation}
    
The gap equations~(\ref{gap1})-(\ref{gap4})  for the condensates are equivalent to the extremum condition of the thermodynamic potential
\begin{equation}
    \frac{\partial\Omega}{\partial\sigma_q}=\frac{\partial\Omega}{\partial\eta_q}=\frac{\partial\Omega}{\partial\pi}=\frac{\partial\Omega}{\partial\delta}=0.
\end{equation}
The flavor number densities Eq.~(\ref{density}) obtained from the matrix elements of the quark propagator are equivalent to the thermodynamical relations
\begin{equation}
    \rho_q=-\frac{\partial \Omega}{\partial \mu_q}.
\end{equation}
\section{Zero temperature}
\label{sec:results1}
Because of the fact that we simplified the interactions as four-fermion contact pointlike interactions in the Lagrangian, the NJL model cannot be renormalized. To solve the gap equations and calculate the thermodynamic functions numerically, we should first fix the model parameters. It is necessary to introduce a regulator $\Lambda$ that serves as an energy scale at which the strong interaction vanishes, and which can be thought of as indicating the onset of asymptotic freedom \cite{PhysRevD.71.116001}. In the following, we take a hard three-momentum cutoff $\Lambda$, and following Refs. \cite{BoerBoomsma2008,BoerBoomsma2009}, we introduce $c$ as $G=(1-c)G_s$ and $G'=-K/2=cG_s$, the parameter $c$ connecting the two couplings determines the strength of the $\theta$ interaction, Often, in recent studies involving flavor mixing within the NJL model and its extensions, the couplings $G$ and $G'$ are also taken equal. Although the exact value of $c$ is unknown, the splitting of $\eta-\eta'$ in the three-flavor model shows that $c\sim0.2$ is favorable \cite{Frank2003}. The value $c = 0.2$ has also been used in refs. \cite{BoerBoomsma2008,BoerBoomsma2009,SakaiKounoSasakiYahiro2011,PhysRevD.100.014013,PhysRevD.110.014042,PhysRevD.100.076021}. Therefore, we also take $c = 0.2$ here. Hence, the 2-flavor NJL model has three parameters of $m$, $\Lambda$, $G_s$. In this work, we employ parameters $m=0.005 $ $\text{GeV}$, $ \Lambda = 0.66$ $\text{GeV}$, $ G_s = 4.8$ $\text{GeV}^{-2}$, which are fitted by pion mass, decay constant, and the isospin density from a recent lattice QCD study \cite{XU2021115540}.

In this section, we concentrate on the $\theta$ effects with finite isospin chemical potential when both baryon chemical potential and temperature are vanishing. For a given isospin chemical potential $\mu_I$, temperature $T$,  baryon chemical potential $\mu_B$ and a CP-violating parameter $\theta$, we can solve the coupled self-consistent gap equations~(\ref{gap1})–(\ref{gap4}). And we check if the solution corresponds to global minima of thermodynamic potential. Only the one with minimal thermodynamic potential will be chosen.

\begin{figure*}[htbp]
    \centering
    \begin{subfigure}[b]{0.45\textwidth}
        \includegraphics[width=\textwidth]{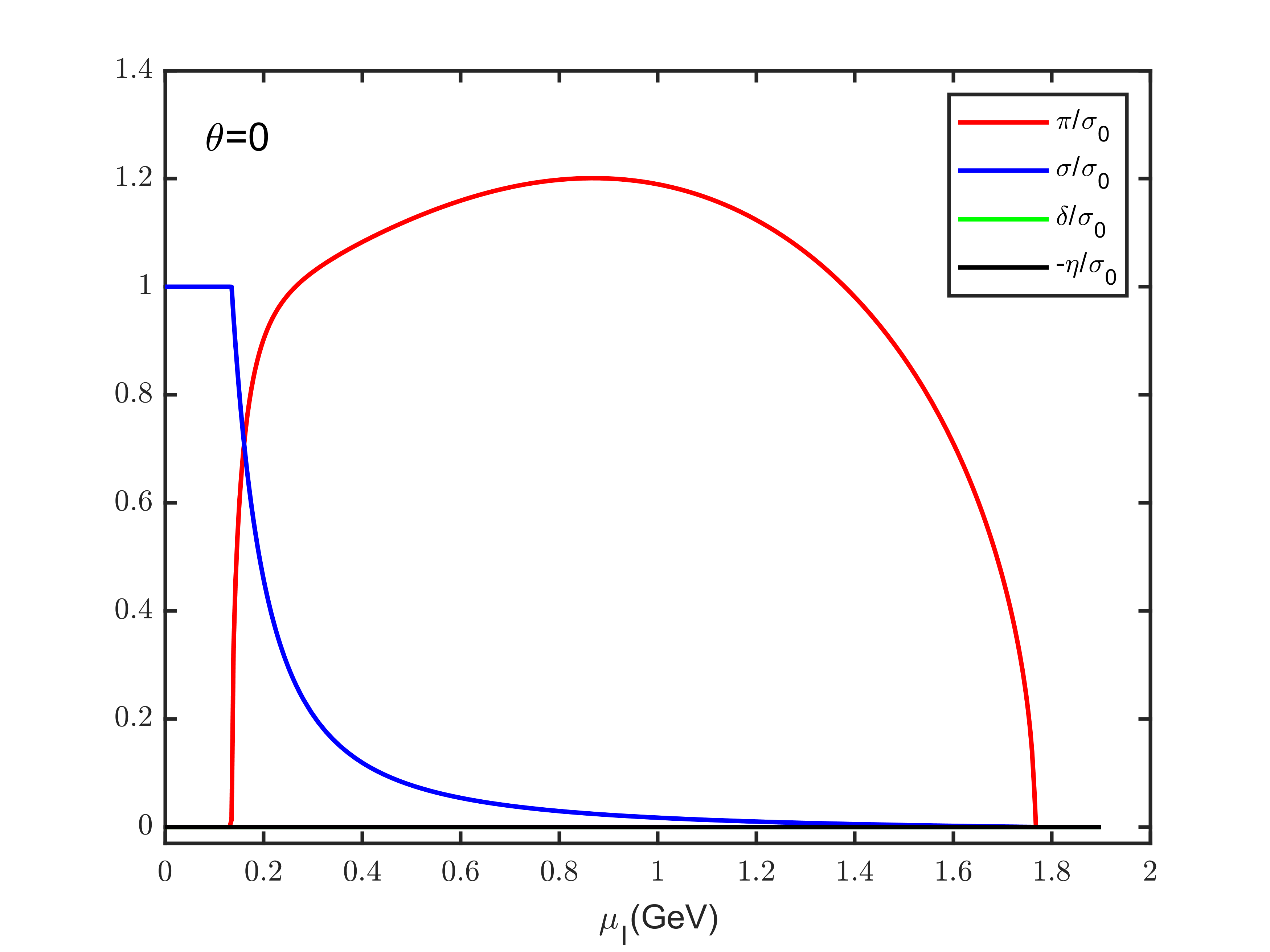}
    \end{subfigure}
    \hspace{-0.93cm} 
    \begin{subfigure}[b]{0.45\textwidth}
        \includegraphics[width=\textwidth]{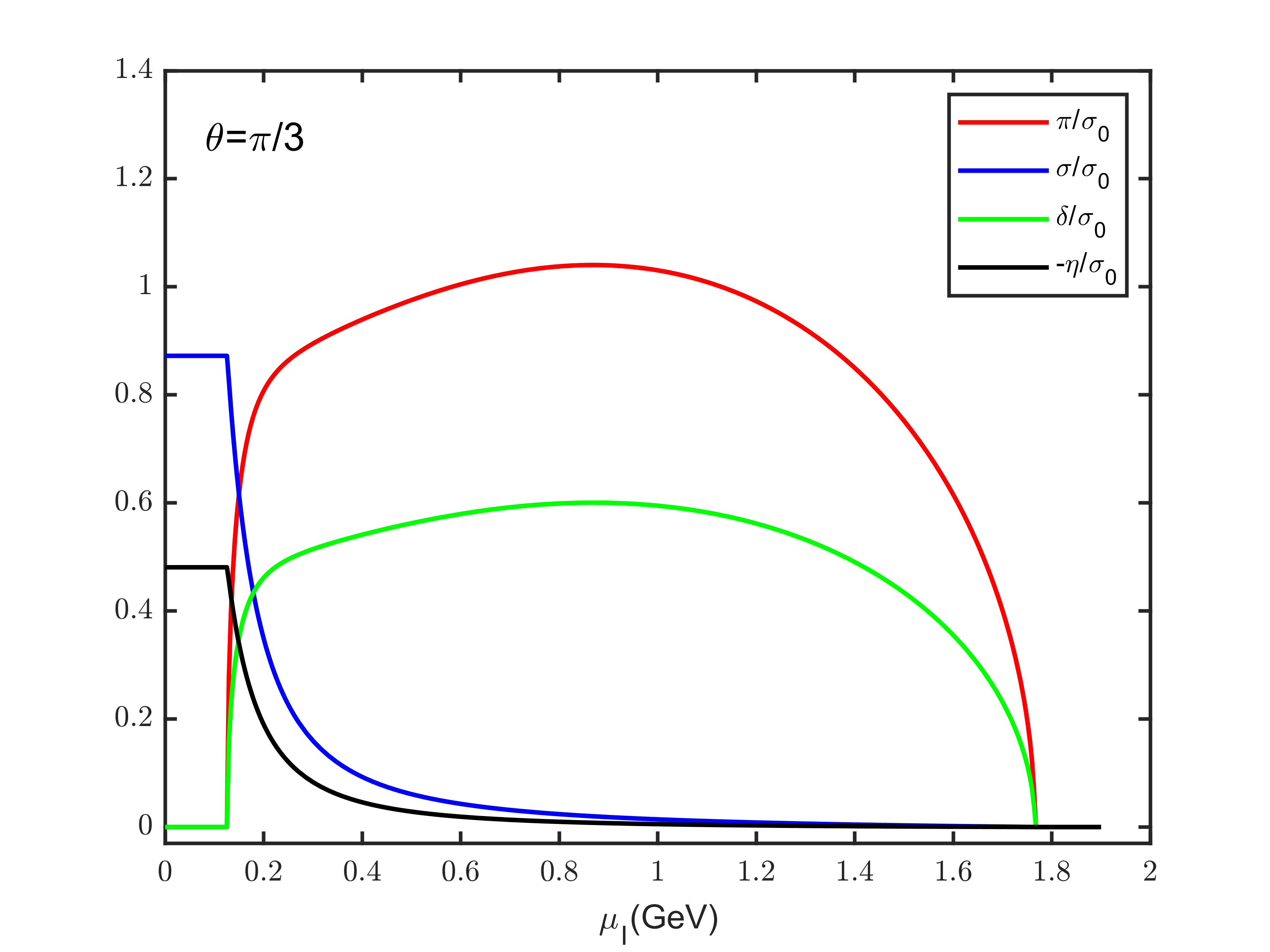}
    \end{subfigure}

    \vspace{0cm} 

    \begin{subfigure}[b]{0.45\textwidth}
        \includegraphics[width=\textwidth]{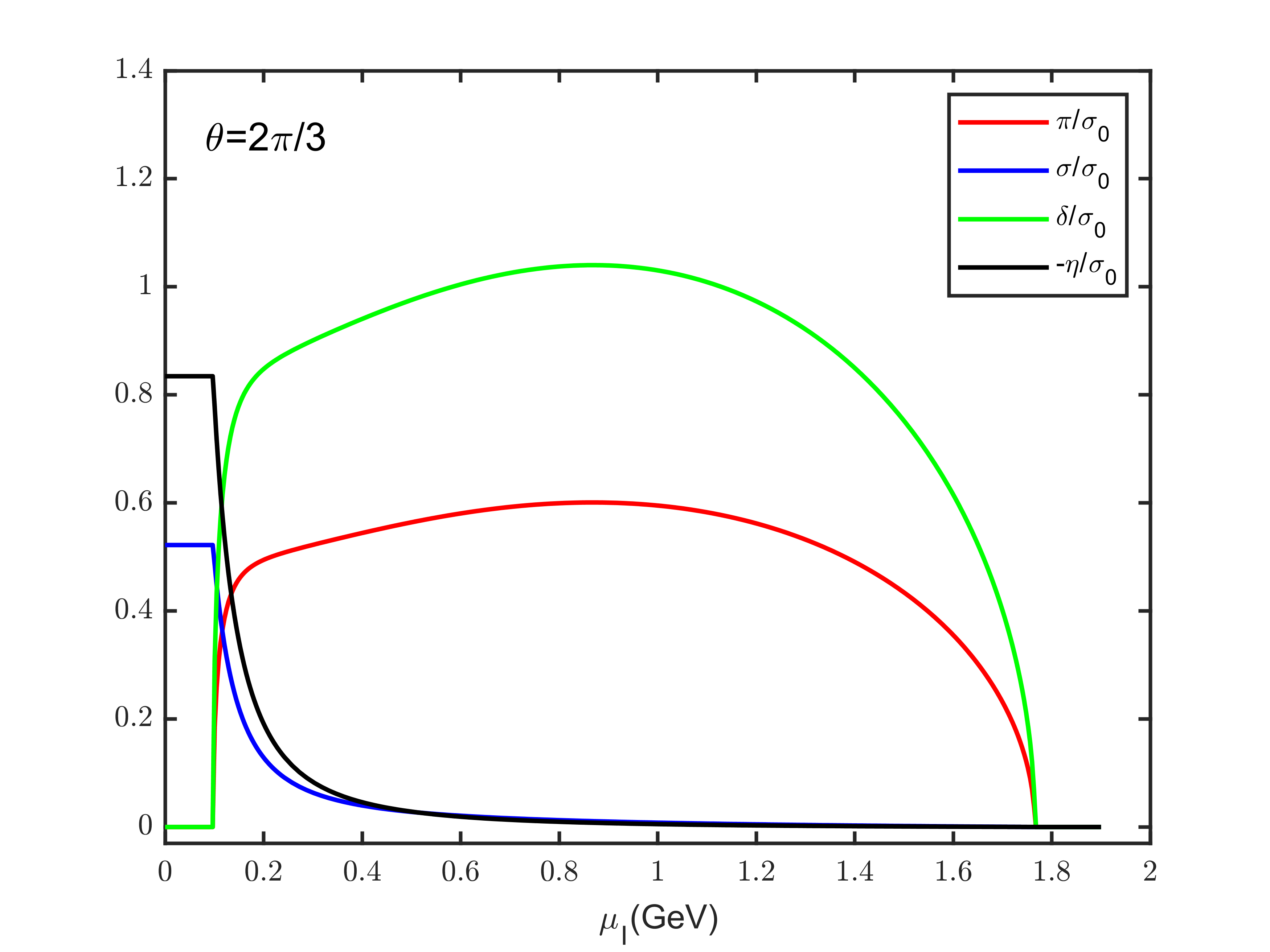}
    \end{subfigure}
    \hspace{-0.93cm} 
    \begin{subfigure}[b]{0.45\textwidth}
        \includegraphics[width=\textwidth]{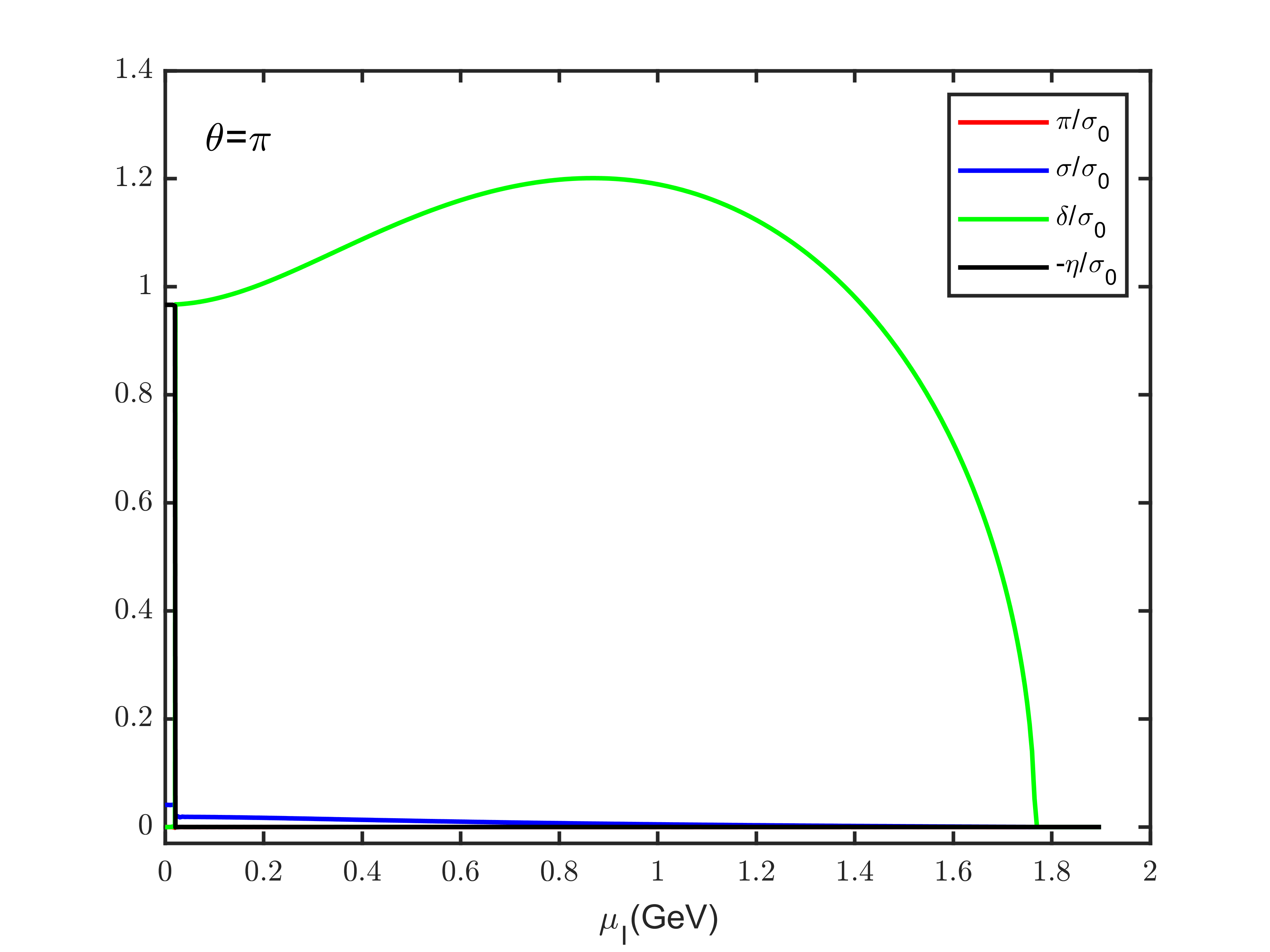}
    \end{subfigure}

    \caption{The four condensates $\sigma$, $\pi$, $\eta$ and $\delta$ scaled by the
chiral condensate $\sigma_0$, as function of isospin chemical potential $\mu_I$ for several values of $\theta \equiv a/f_a$. Top left plot corresponds to $\theta=0$, top right to $\theta=\pi/3$, bottom left to $\theta=2\pi/3$, and finally bottom right to $\theta=\pi$.}
    \label{fig1}
\end{figure*}

In Fig.~\ref{fig1}, we show the condensates as a function of isospin chemical potential for $\theta=0$, $\pi/3$, $2\pi/3$, and $\pi$. We have normalized the quark condensates with chiral condensate $\sigma_0$ in the vacuum. In Fig.~\ref{fig1} ($\theta=0$), we see that the condensates $\eta$ and $\delta$ are both zero, consistent with the results of standard NJL model studies. It is well known that in the isospin symmetric phase, the vacuum state is not disturbed by small isospin chemical potential so that chiral condensate keeps its vacuum value and $\pi=0$. At the critical isospin chemical potential $\mu_I^{\text{crit}}=0.135$ GeV (which means it exceeds the pion mass \cite{HE200593}), the isospin symmetry starts to break spontaneously and the pion condensate appears. Both the chiral and pion superfluidity phase transitions are of second order. The pion condensate increases with $\mu_I$ and at the same time the chiral condensate decreases with $\mu_I$. The two condensates coexist in a wide region. The reason for the disappearance of pion condensation at high $\mu_I$ can be explained by asymptotic freedom of QCD.

In Fig.~\ref{fig1} ($\theta=\pi/3$), the pseudoscalar condensate $\eta$ has a nonvanishing value in the isospin symmetric phase with nonvanishing $\theta$. This matches the conclusions in Refs.~\cite{PhysRevD.85.114008}. At the critical isospin chemical potential $\mu_I^{\text{crit}}=0.126$ GeV (lower than the $\theta=0$ case), the isospin symmetry starts to break spontaneously with simultaneous emergence of both the pseudoscalar condensate $\pi$ and scalar condensate $\delta$ through second-order phase transitions. The pion condensate and $\delta$ condensate increase with $\mu_I$, while the chiral condensate and $\eta$ condensate decrease with $\mu_I$. All four condensates coexist across a broad $\mu_I$ range. Importantly, both $\pi$ and $\delta$ condensates vanish simultaneously at high $\mu_I$, demonstrating that spontaneous isospin symmetry breaking under nonzero $\theta$ is jointly manifested by the pion condensate and $\delta$ condensate. Throughout this process, we observe nearly identical functional dependencies: the pion condensate and $\delta$ condensate exhibit similar growth rates, while the chiral condensate and $\eta$ condensate show analogous decay patterns. The sole distinction lies in the relative magnitudes: normal condensates ($\sigma$, $\pi$) maintain magnitudes larger than their counterparts ($\eta$, $\delta$).

In Fig.~\ref{fig1} ($\theta=2\pi/3$), the condensates $\sigma$ and $\eta$ display complementary $\theta$ dependence in the symmetric isospin phase, preserving an approximately constant total condensate $\sqrt{\sigma^2+\eta^2}$ . In the broken phase with reduced critical $\mu_I^{\text{crit}}=0.097$ GeV, the condensates have an analogous behavior as before only the complementary behavior reverses: the magnitudes of ($\sigma$, $\pi$) become smaller than those of ($\eta$, $\delta$). Analogous complementarity emerges between the $\pi$ and $\delta$ condensates as $\theta$ increases in the isospin symmetry-broken phase.

In Fig.~\ref{fig1} ($\theta=\pi$), the scalar condensate $\sigma$ nearly vanishes (retaining a small residual magnitude from nonzero current quark masses) in the isospin symmetric phase. The isospin symmetry breaking transition becomes first order at $\mu_I^{\text{crit}}=0.021$ GeV, contrasting with second order transitions at other $\theta$ values. This transition order difference aligns with expectations, as the transition for the condensate variations with temperature becomes a second order transition at $\theta=\pi$ instead of a crossover for lower $\theta$~\cite{PhysRevD.85.114008}. The condensates of $\eta$ and pion completely vanish in the isospin symmetry breaking phase and the vanishing of the $\eta$ condensate signals the restoration of CP symmetry which means that spontaneous breaking of isospin symmetry occurs simultaneously with the restoration of  CP symmetry and the phase of isospin symmetry breaking is also the CP symmetry phase.  The scalar condensate $\sigma$ persists with a minimal value due to explicit chiral symmetry breaking, while the $\delta$ condensate dominates the isospin-symmetry-broken phase. Last but not least, all condensates ultimately vanish at high $\mu_I$ values related to the finite momentum cutoff $\Lambda$ in the NJL framework, regardless of $\theta$.

We can preliminarily conclude that $\theta$ has two effects: suppressing normal condensates $\sigma$ and $\pi$, promoting condensates $\eta$ and $\delta$, and catalyzing spontaneous isospin symmetry breaking. The striking similarity between ($\sigma$, $\pi$) and ($\eta$, $\delta$) evolution patterns will be further verified under conditions of finite temperature and baryon density.

Next we consider the $\theta$ dependence of the ground state structure with a pure isospin effect. 

\begin{figure}[htbp]
    \centering
    \includegraphics[width=8.6cm]{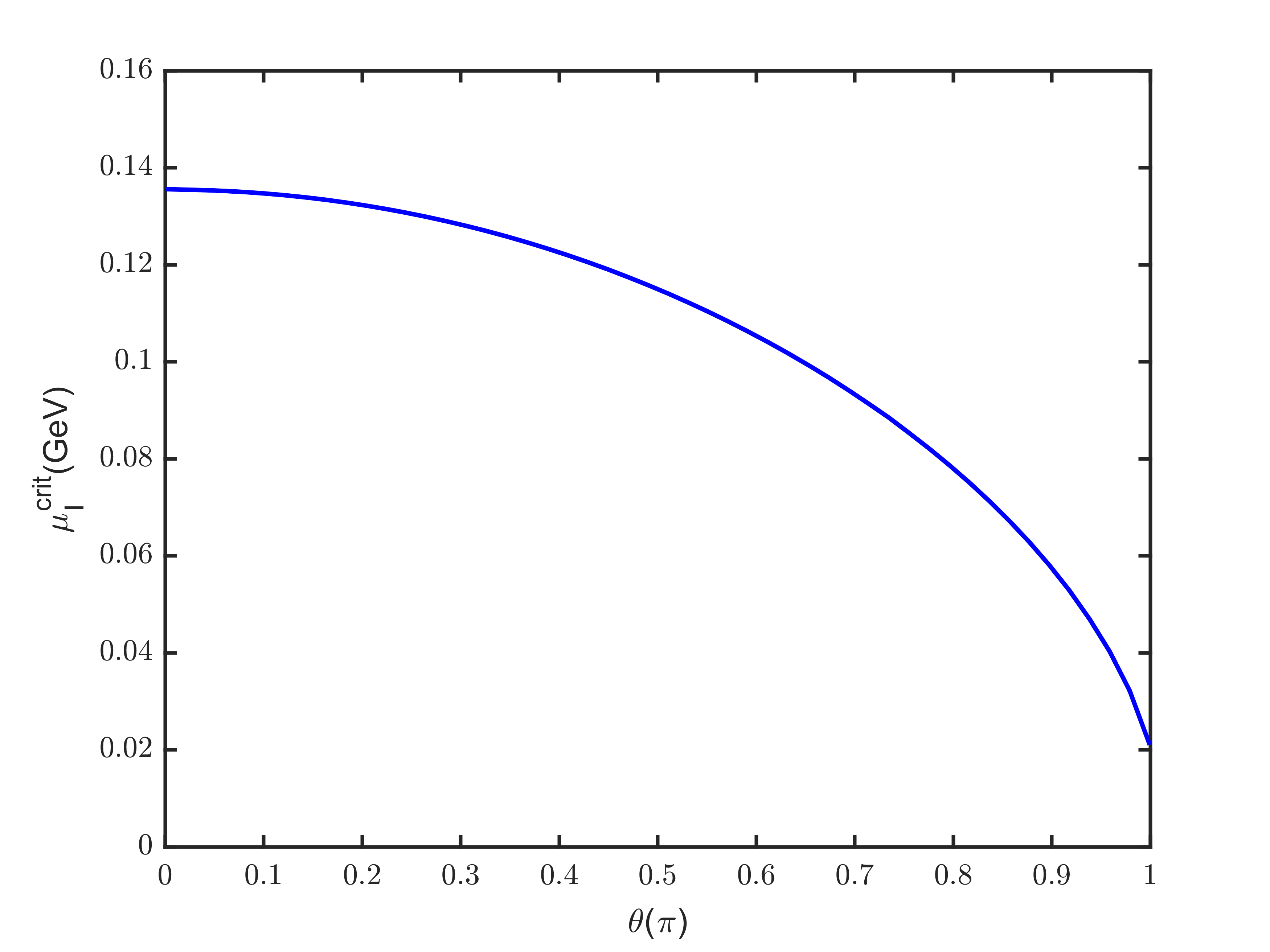}
    \caption{The critical isospin chemical potential for spontaneous isospin symmetry breaking, $\mu_{I}^{\text{crit}}$ as function of CP-violating parameter $\theta$.}
    \label{fig2a}
\end{figure}

In Fig.~\ref{fig2a}, we first show the $\theta$ dependence of the critical isospin chemical potential for spontaneous isospin symmetry breaking, $\mu_{I}^{\text{crit}}(\theta)$. In particular, the second-order isospin phase transition exhibits a strong $\theta$ sensitivity. We observe that the critical isospin chemical potential decreases with increasing $\theta$, with the descending rate accelerating at larger $\theta$ values. It should be noted that in previous studies \cite{PhysRevD.111.L051501}, increasing $\theta$ was also found to reduce the baryon chemical potential threshold for first-order phase transitions. This suggests that larger $\theta$ values induce system instability, thus catalyzing phase transitions at lower critical thresholds.

\begin{figure}[htbp]
    \centering
    \includegraphics[width=8.6cm]{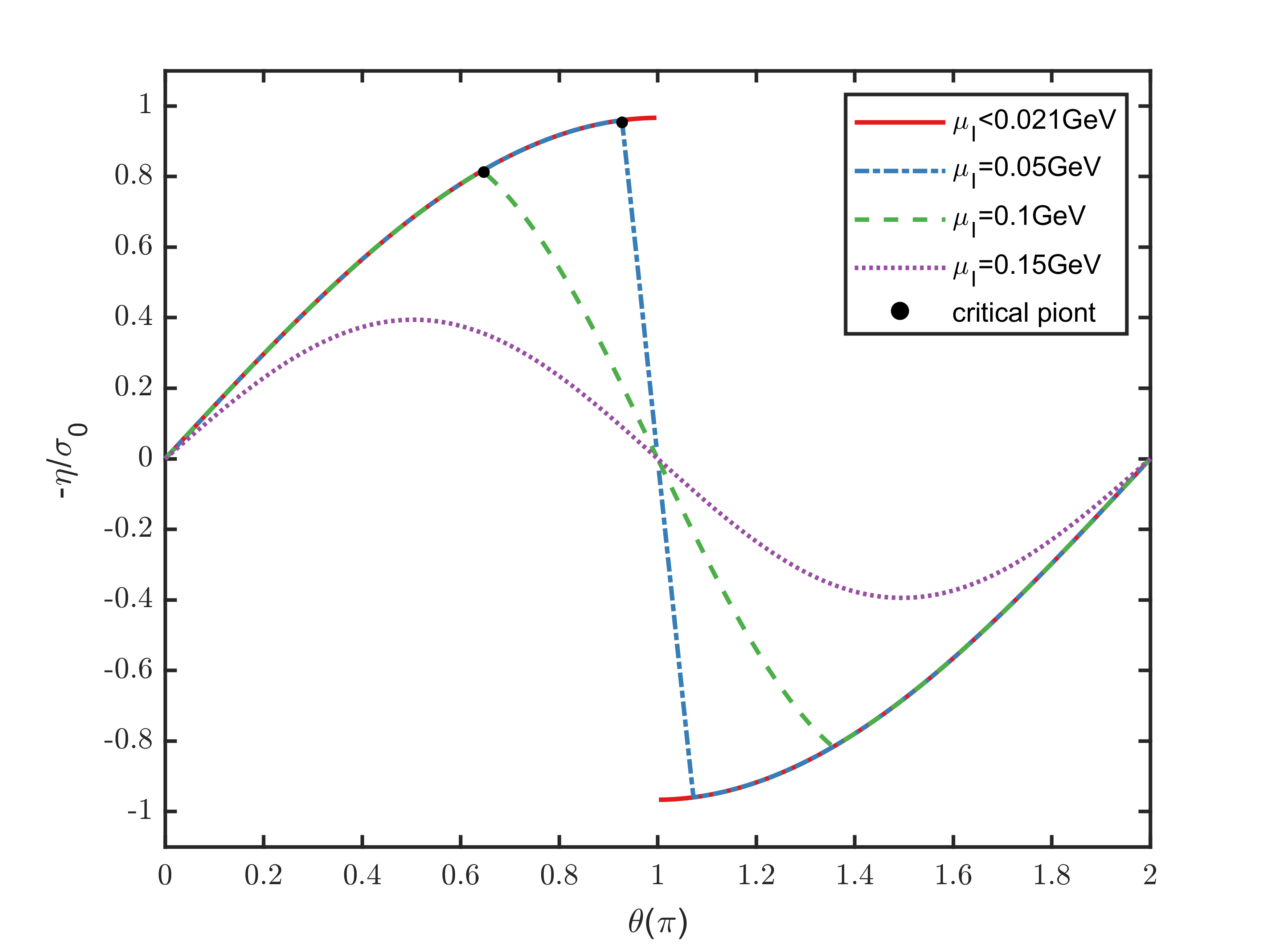}
    \caption{The variations of the pseudoscalar condensate $\eta$ with respect to $\theta$ for several isospin chemical potentials. The black dots are the positions of the critical $\theta_c$ when spontaneous isospin symmetry breaking occurs.}
    \label{fig2b}
\end{figure}
In Fig.~\ref{fig2b}, we show the variations of the pseudoscalar condensate $\eta$ with respect to $\theta$ (or equivalently the scaled axion field $a/f_a$) for several isospin chemical potentials, considering both subcritical and supercritical values relative to the CP symmetry restoration threshold. The pseudoscalar condensate $\eta$ exhibits a periodic $\theta$-dependence with a $2\pi$ periodicity. 
For $\mu_I < 0.021$ GeV, spontaneous CP violation manifests itself clearly in $\theta=\pi+2k\pi$, where $\eta$ bifurcates into two degenerate solutions differing only in sign. Throughout this range $\mu_I$, the system remains in the isospin-symmetric phase without spontaneous symmetry breaking for any $\theta$.
In the intermediate regime $0.021\,\text{GeV} \leq \mu_I < 0.135\,\text{GeV}$, isospin symmetry remains unbroken in the $\theta=0$ vacuum. However, at a critical $\theta_c(\mu_I)$ (marked by black dots in the figure), spontaneous isospin symmetry breaks. For $\theta < \theta_c$, the system persists in the symmetric phase with $\eta$ maintaining its $\theta$ dependent equilibrium value. When $\theta > \theta_c$ (i.e. $\mu_I > \mu_{I}^\text{crit}(\theta)$), $\eta$ decreases monotonically and eventually vanishes at $\theta=\pi$. The $\eta$-degeneracy at $\theta=\pi$ observed at lower $\mu_I$ is lifted beyond $\mu_{I}^\text{crit}=0.021$ GeV (see Fig.~\ref{fig1}), signaling the breakdown of Dashen's phenomenon.
For $\mu_I > 0.135$ GeV, spontaneous isospin symmetry breaking already exists in the vacuum $\theta=0$, eliminating any $\theta$ interval where $\eta$ remains static. Across all $\theta$, $\eta$ evolves under competing effects: (1) $\theta$-enhancement of  $\eta$ pseudoscalar condensation; (2) suppression from acilitating spontaneous isospin symmetry breaking. The enhancement driven by $\theta$ dominates at low $\theta$, increasing $\eta$, while the suppression of symmetry breaking prevails at high $\theta$, causing the reduction $\eta$.

\begin{figure}[htbp]
    \centering
    \includegraphics[width=8.6cm]{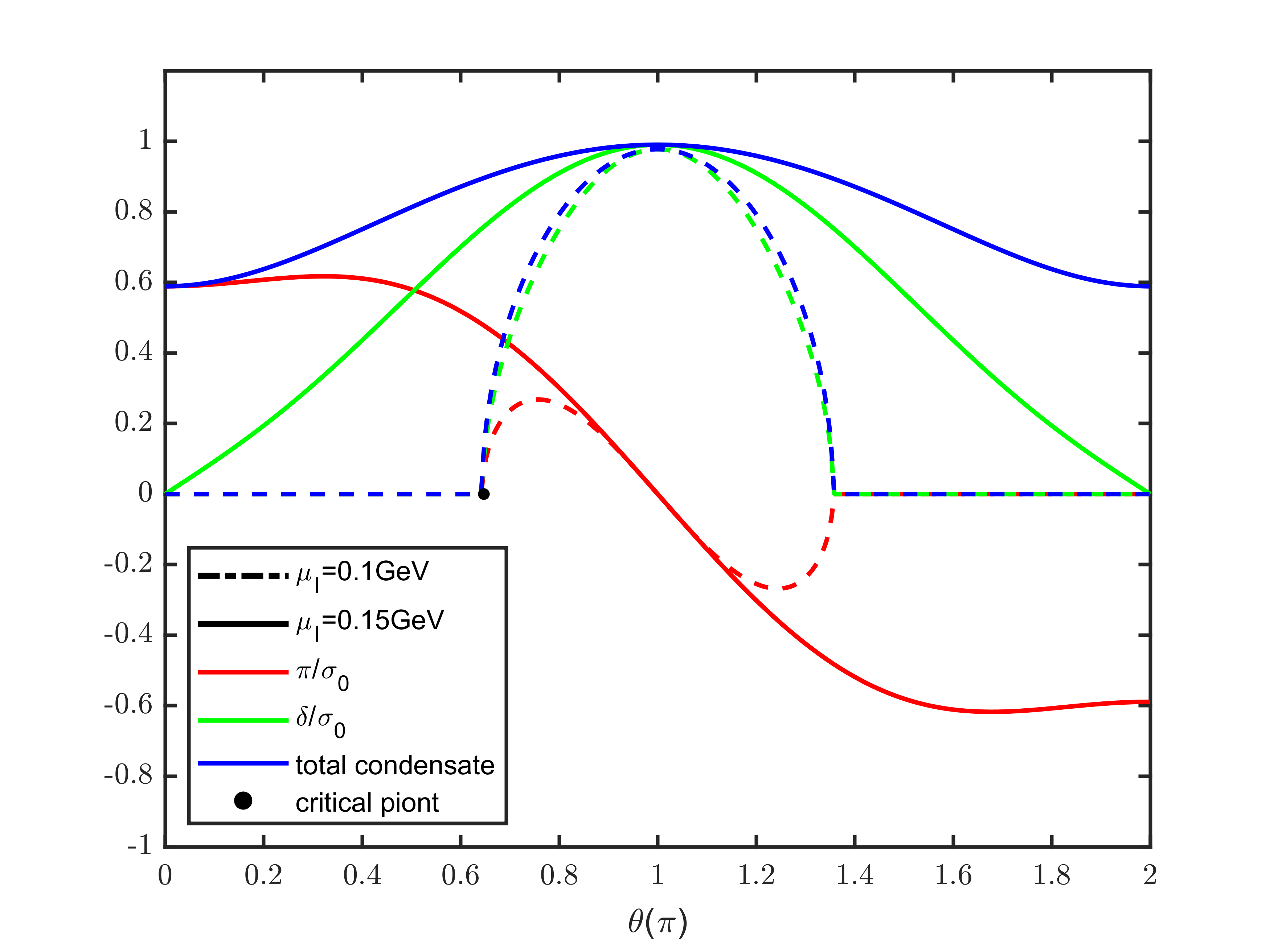}
    \caption{The condensates $\pi$ and $\delta$ as functions of $\theta$  for several isospin chemical potentials. The black dots are the positions of the critical $\theta_c$ when spontaneous isospin symmetry breaking occurs.}
    \label{fig2c}
\end{figure}

In Fig.~\ref{fig2c}, we present the $\theta$ dependence of pseudoscalar condensate $\pi$ and scalar condensate $\delta$. As established in previous studies \cite{PhysRevD.85.114008}, the $\sigma$-$\eta$ condensate pair exhibits a complementary relationship with approximately conserved total condensate $\sqrt{\sigma^2+\eta^2} \approx \text{const}$. We investigate whether this behavior extends to the $\pi$-$\delta$ pair in the isospin-symmetry-broken phase.

For $\mu_I$ slightly above 0.135 GeV (where pion condensation exists at $\theta=0$), the $\pi$-$\delta$ system shows \textit{approximate} complementarity. The decreasing rate of $\pi$ with $\theta$ remains slower than the increasing rate of $\delta$, leading to a monotonic increase in total condensate $\sqrt{\pi^2+\delta^2}$. This behavior stems from competing $\theta$-effects: 
1. Suppression of $\pi$; 
2. Enhancement of $\delta$; 
3. catalyzing isospin symmetry breaking.  
These mechanisms accelerate $\delta$ growth while decelerating $\pi$ suppression, collectively increasing total condensate. 
In the range $0.021\,\text{GeV} \leq \mu_I < 0.135\,\text{GeV}$, isospin symmetry remains intact at $\theta=0$ but spontaneously breaks at critical $\theta_c(\mu_I)$ (marked by a black dot). For $\theta < \theta_c$ (isospin-symmetric phase), both $\pi$ and $\delta$ vanish. When $\theta > \theta_c$ ($\mu_I > \mu_{I}^\text{crit}(\theta)$):
$\delta$ increases monotonically; $\pi$ first rises, then falls, vanishing at $\theta=\pi$. 
At $\theta=\pi$, both condensates exhibit continuity throughout the transition. For $\theta > \pi$, their signs are reversed. The total condensate $\sqrt{\pi^2+\delta^2}$ remains non-zero throughout the symmetry-broken phase.
For $\mu_I < 0.021\,\text{GeV}$, the system persists in the isospin symmetric phase with $\sqrt{\pi^2+\delta^2}=0$ across all $\theta$, as spontaneous symmetry breaking never occurs.

\begin{figure}[htbp]
    \centering
    \includegraphics[width=8.6cm]{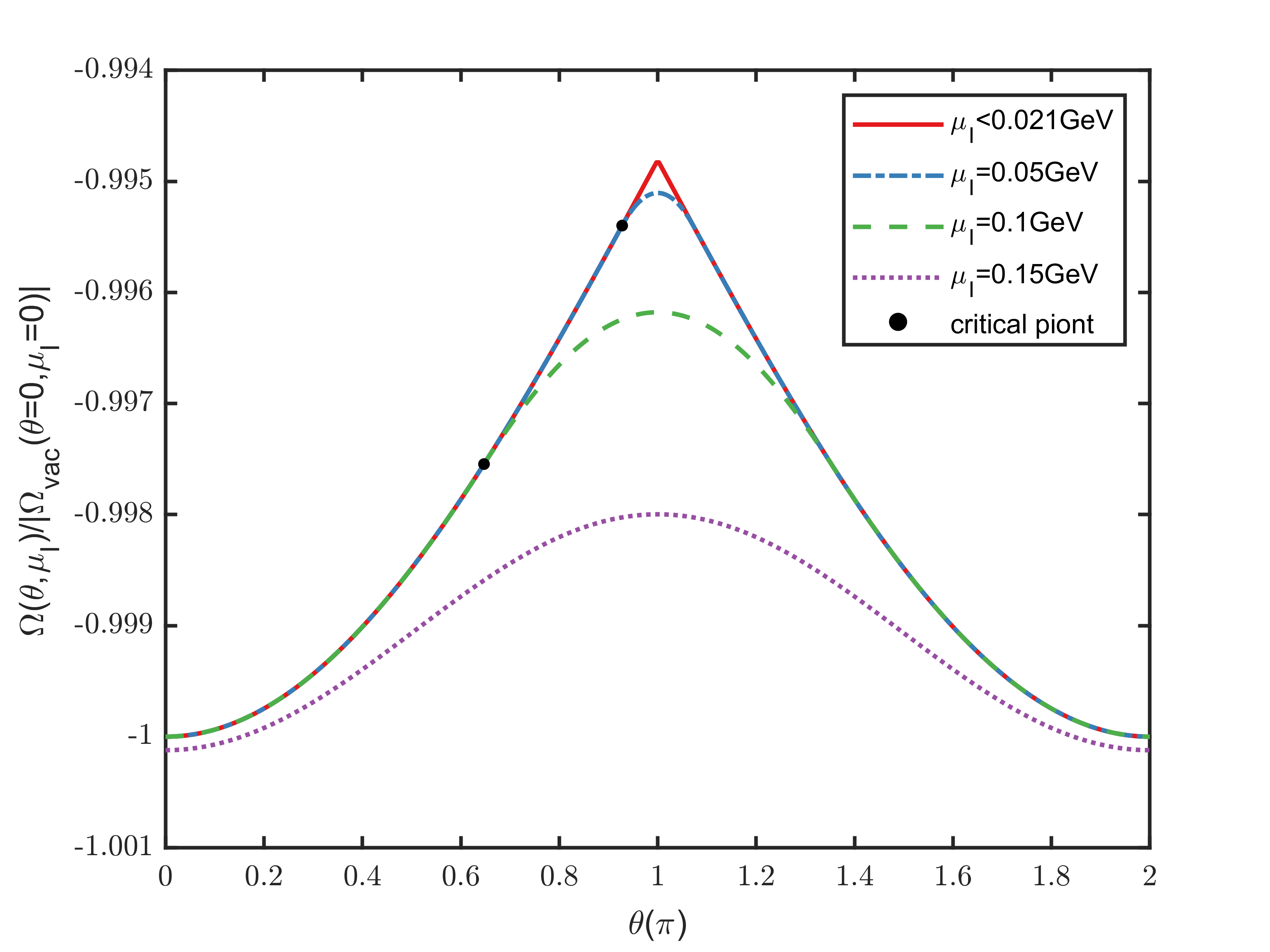}
    \caption{The normalized thermodynamic potential as a function of $\theta$ for different isospin chemical potentials. The black dots are the positions of the critical $\theta_c$ when spontaneous isospin symmetry breaking occurs.}
    \label{fig2d}
\end{figure}
In Fig.~\ref{fig2d}, we display the normalized thermodynamic potential (axion potential) as a function of $\theta$ for different isospin chemical potentials $\mu_I$. The potential minimum resides at $\theta=2i\pi$, consistent with the Vafa-Witten theorem, while maxima occur at $\theta=(2i+1)\pi$ - a feature also observed in NJL model studies \cite{PhysRevD.103.074003,PhysRevD.100.014013}. 
For $\mu_I < 0.021\,\mathrm{GeV}$, the system remains in the isospin-symmetric phase across all $\theta$, with the effective potential not affected by $\mu_I$. Spontaneous CP violation at $\theta=\pi$ induces a discontinuity in the potential's first derivative at this point.
As $\mu_I$ increases from $0.021\,\mathrm{GeV}$ to $0.135\,\mathrm{GeV}$, the potential profile remains unchanged for $\theta < \theta_c$ but becomes flattened for $\theta > \theta_c$. This behavior reflects the absence of spontaneous isospin symmetry breaking below $\theta_c$ and its activation above $\theta_c$. The restoration of CP symmetry at $\theta=\pi$ restores derivative continuity, producing smooth potential transitions.
When $\mu_I$ exceeds $0.135\,\mathrm{GeV}$ (where spontaneous isospin breaking exists at $\theta=0$), two key modifications emerge: 
1. The value of the thermodynamic potential at $\theta=0$ is smaller; 
2. The potential landscape flattens significantly throughout the $\theta$ domain. 

\section{finite temperature}
\label{sec:results2}
We now study the temperature behavior of condensates and thermodynamic functions in the phase of isospin symmetry breaking with $\theta$ effects, and discuss the isospin symmetry breaking phase diagram in the $T$-$\mu_I$ plane at $\mu_B=0$. Then we extend the phase diagram for the CP transition in the $T$-$\mu_I$ plane, which was discussed in the $T$-$\mu_B$ plane in Ref.~\cite{PhysRevD.85.114008}.

\begin{figure*}[htbp]
    \centering
    \begin{subfigure}[b]{0.45\textwidth}
        \includegraphics[width=\textwidth]{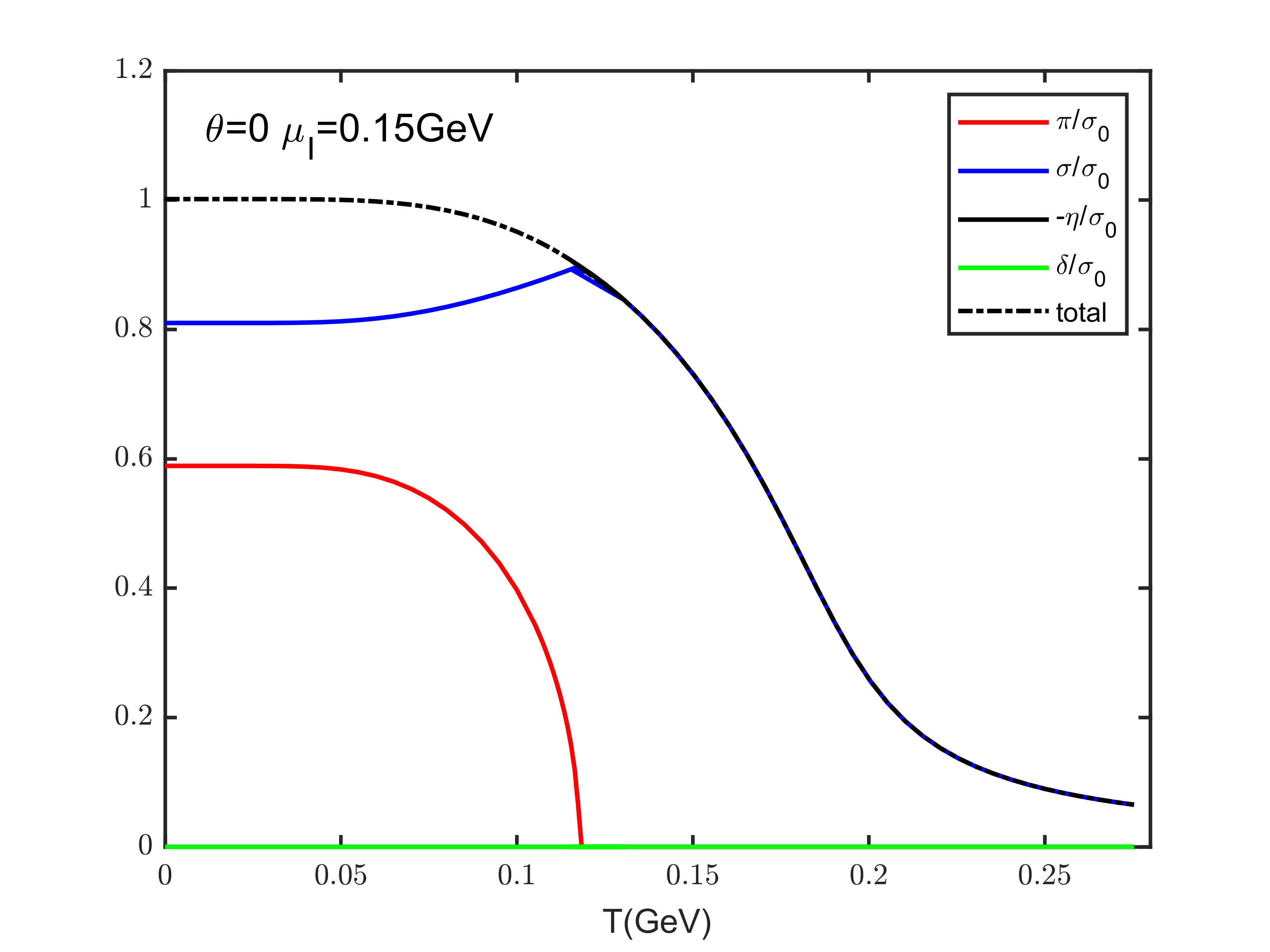}
    \end{subfigure}
    \hspace{-0.93cm} 
    \begin{subfigure}[b]{0.45\textwidth}
        \includegraphics[width=\textwidth]{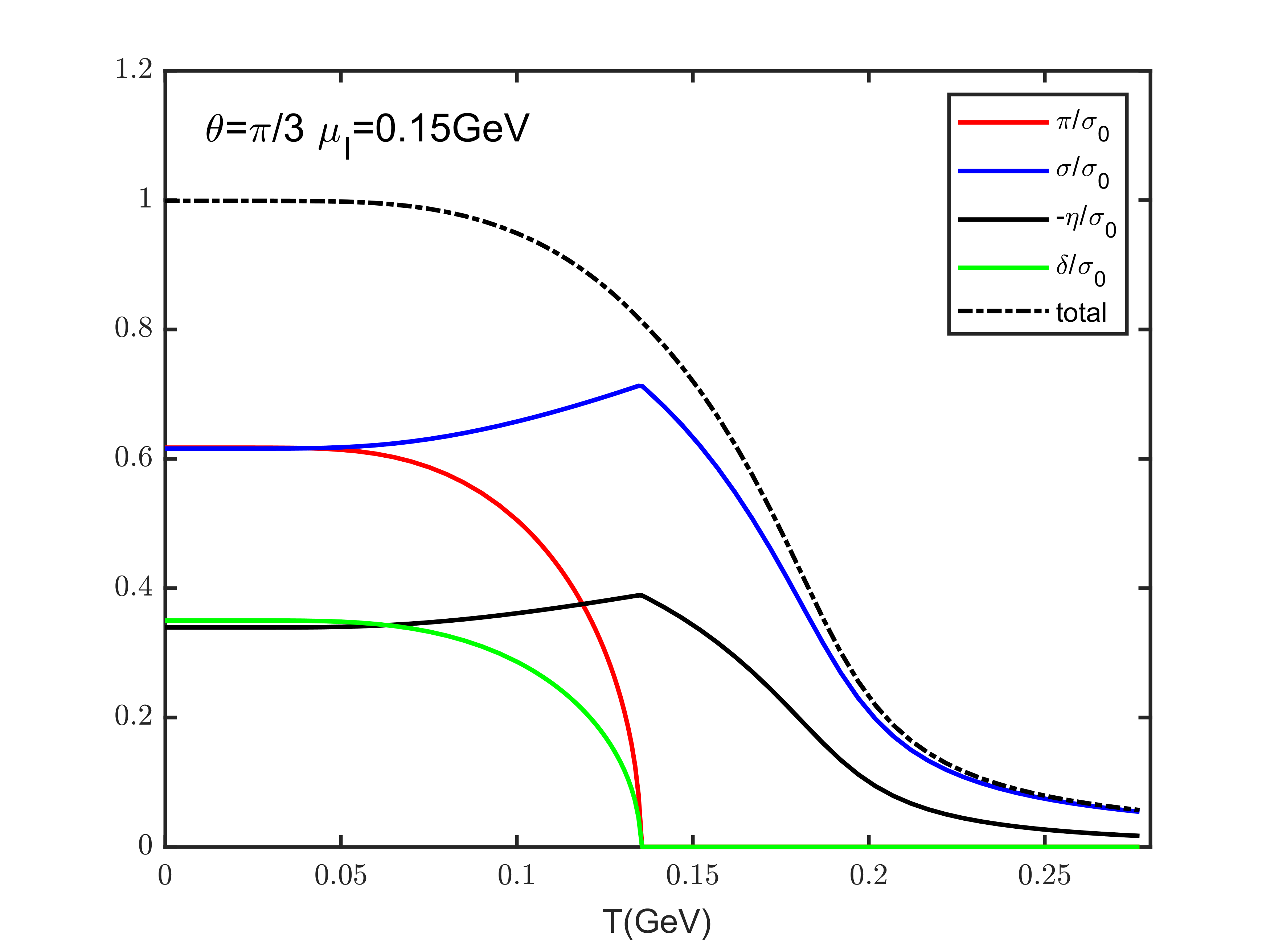}
    \end{subfigure}

    \vspace{0cm} 

    \begin{subfigure}[b]{0.45\textwidth}
        \includegraphics[width=\textwidth]{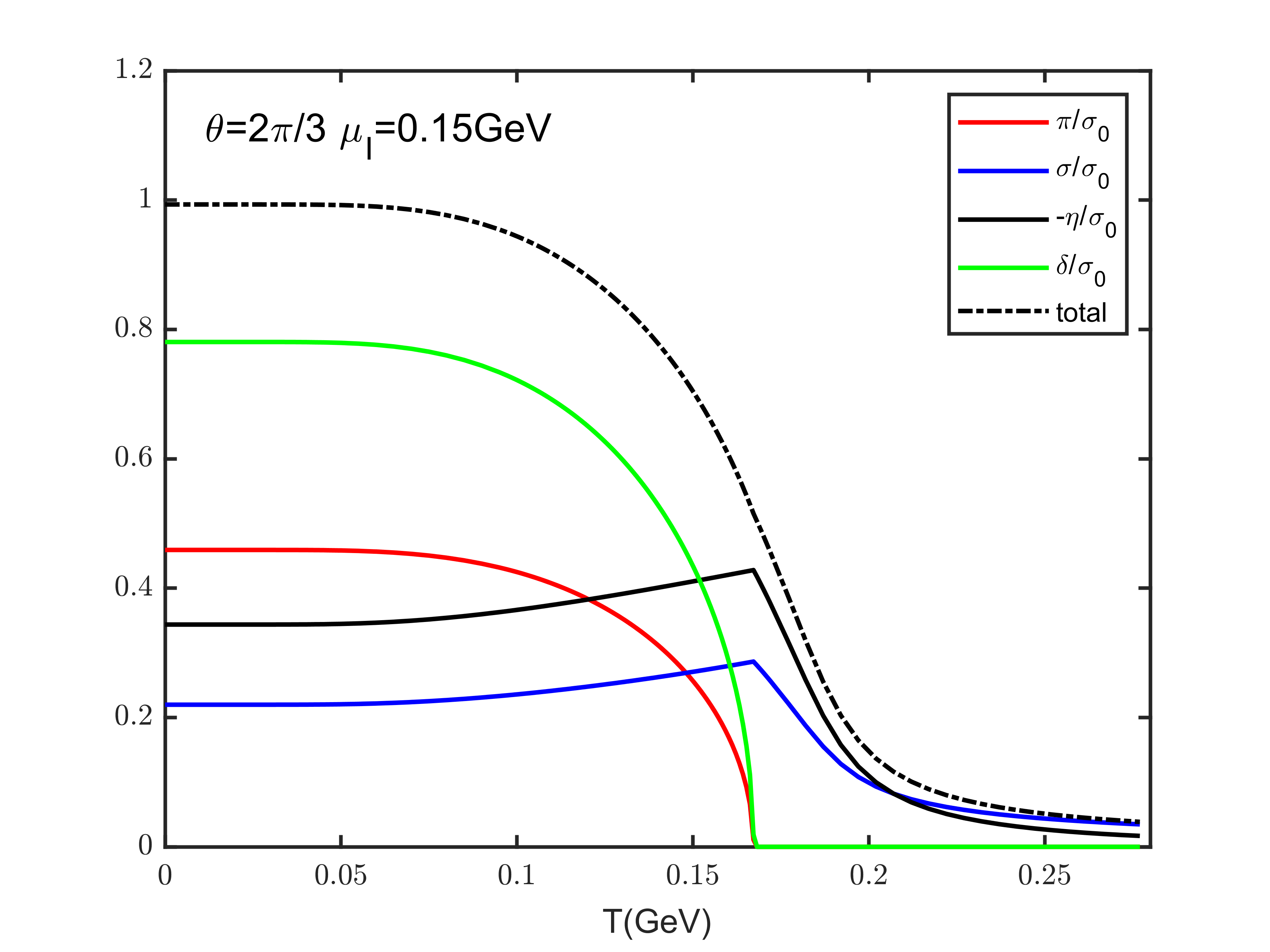}
    \end{subfigure}
    \hspace{-0.93cm} 
    \begin{subfigure}[b]{0.45\textwidth}
        \includegraphics[width=\textwidth]{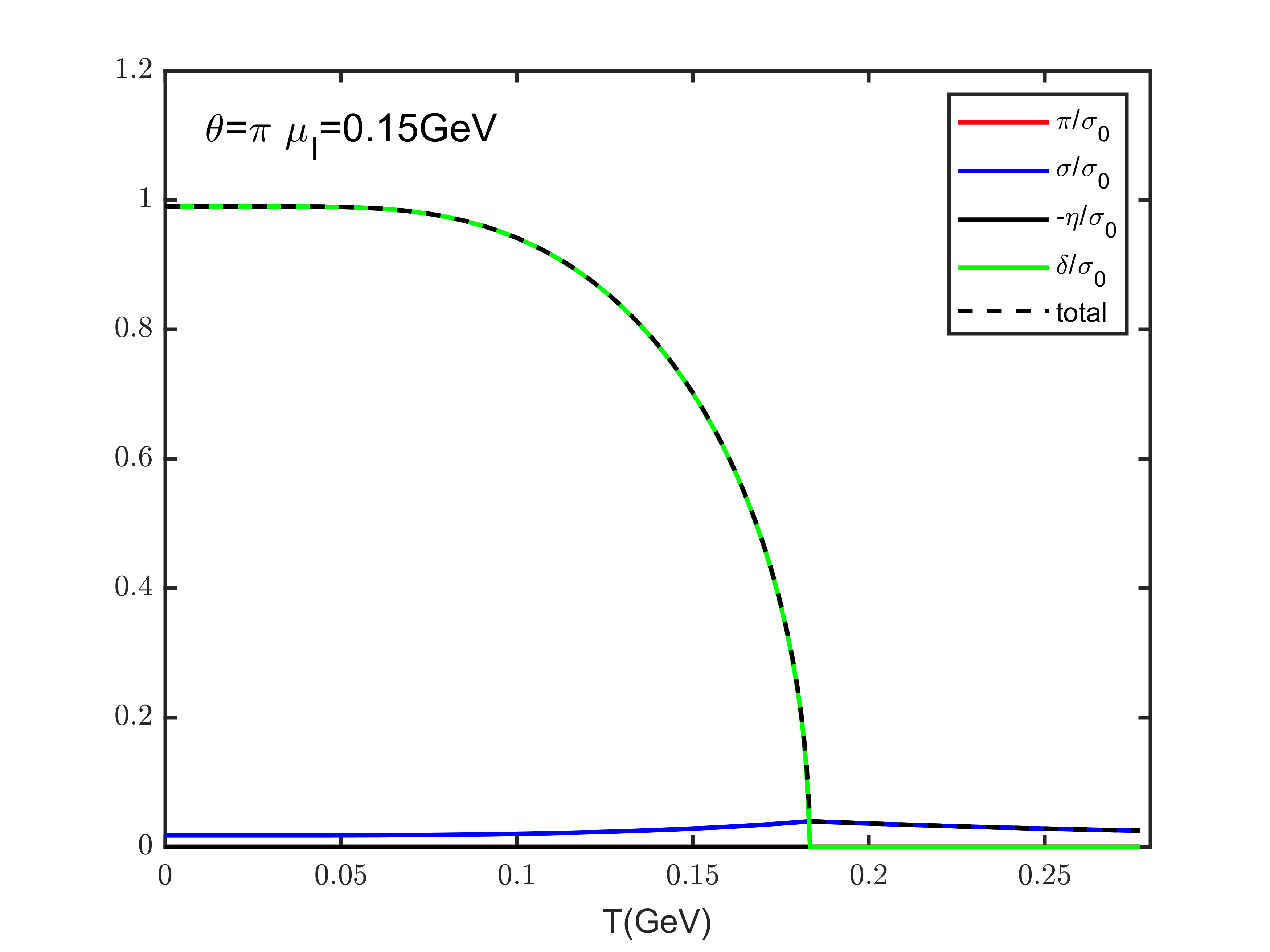}
    \end{subfigure}

    \caption{The four condensates $\sigma$, $\pi$, $\eta$ and $\delta$ scaled by the
chiral condensate $\sigma_0$, as function of temperature $T$ at $\mu_B=0$ and $\mu_I=0.15$ GeV for several values of $\theta \equiv a/f_a$. The dashed lines
are for the total condensate $\sqrt{\sigma^2+\pi^2+\eta^2+\delta^2}/|\sigma_0|$. Top left plot corresponds to $\theta=0$, top right to $\theta=\pi/3$, bottom left to $\theta=2\pi/3$, and finally bottom right to $\theta=\pi$.}
    \label{fig3}
\end{figure*}
In Fig.~\ref{fig3}, we present the temperature dependence of the condensates $\sigma$, $\eta$, $\pi$ and $\delta$ for $\theta=0$, $\pi/3$, $2\pi/3$, and $\pi$ at a fixed isospin chemical potential $\mu_I=0.15$ GeV, maintaining the system in the isospin symmetry-broken phase. All condensates are normalized by the vacuum chiral condensate $\sigma_0$. 

In Fig.~\ref{fig3} ($\theta=0$), the condensates $\eta$ and $\delta$ vanish identically, consistent with the results of the standard NJL model. The pion condensate is already nonzero at the beginning while the strength of $\sigma$ condensate is larger, then the pion condensate decreases due to the temperature effect, while the chiral condensate increases continuously in the coexistence region of the two condensates which is qualitatively different from the decrease of $\sigma$ with increasing temperature in the study on chiral symmetry restoration. The coexistence phase persists until a second-order phase transition restores isospin symmetry at critical temperature $T_c=0.118$ GeV, and $\sigma$ decreases in the region where the isospin symmetry is restored. Notably, the total condensate $\sqrt{\sigma^2+\pi^2}$ mimics the smooth crossover behavior characteristic of chiral phase transitions, despite the distinct microscopic mechanisms governing $\sigma$ and $\pi$ evolution.

In Fig.~\ref{fig3} ($\theta=\pi/3$), as $\theta$ increases, the condensates $\eta$ and $\delta$ begin to acquire nonzero values. Under the $\theta$-enhanced spontaneous isospin symmetry breaking effect, the strengths of $\sqrt{\pi^2+\delta^2}$ and $\sqrt{\sigma^2+\eta^2}$ initially become comparable. Both the pion condensate $\pi$ and $\delta$ condensate decrease with temperature, while the chiral condensate $\sigma$ and $\eta$ condensate increase continuously within the four condensates coexistence region. Eventually, $\pi$ and $\delta$ vanish simultaneously through a second-order phase transition restoring isospin symmetry at a higher critical temperature $T_c = 0.136$ GeV, followed by suppression of $\sigma$ and $\eta$ in the restored phase. Throughout this process, the temperature-dependent behaviors of ($\pi$, $\delta$) and ($\sigma$, $\eta$) remain similar, with the only distinction being that the normal condensates ($\sigma$, $\pi$) maintain larger magnitudes than their counterparts ($\eta$, $\delta$).

In Fig.~\ref{fig3} ($\theta=2\pi/3$), the initial strength of $\sqrt{\pi^2+\delta^2}$ exceeds $\sqrt{\sigma^2+\eta^2}$, and the critical temperature $T_c = 0.167$ GeV for isospin symmetry restoration increases further with $\theta$, both consistent with theoretical expectations. The complementary $\theta$-dependent behaviors between ($\sigma$, $\eta$) and ($\pi$, $\delta$) channels persist, changing the relative hierarchy where condensates ($\eta$, $\delta$) exhibit larger magnitudes than  normal condensates ($\sigma$, $\pi$).

In Fig.~\ref{fig3} ($\theta=\pi$), the $\delta$ condensate still dominates completely in the isospin symmetry-broken phase at finite temperature, while the scalar condensate $\sigma$ retains small nonzero values due to explicit chiral symmetry breaking. Both $\eta$ and $\pi$ condensates remain absent. The $\delta$ condensate vanishes via second-order transition at a larger critical temperature $T_c = 0.183$ GeV. The total condensate $\sqrt{\sigma^2+\eta^2+\pi^2+\delta^2}$ universally displays chiral-crossover-like temperature dependence across all $\theta$ values except $\theta=\pi$.

\begin{figure*}[htbp]
    \centering
    \begin{subfigure}[b]{0.45\textwidth}
        \includegraphics[width=\textwidth]{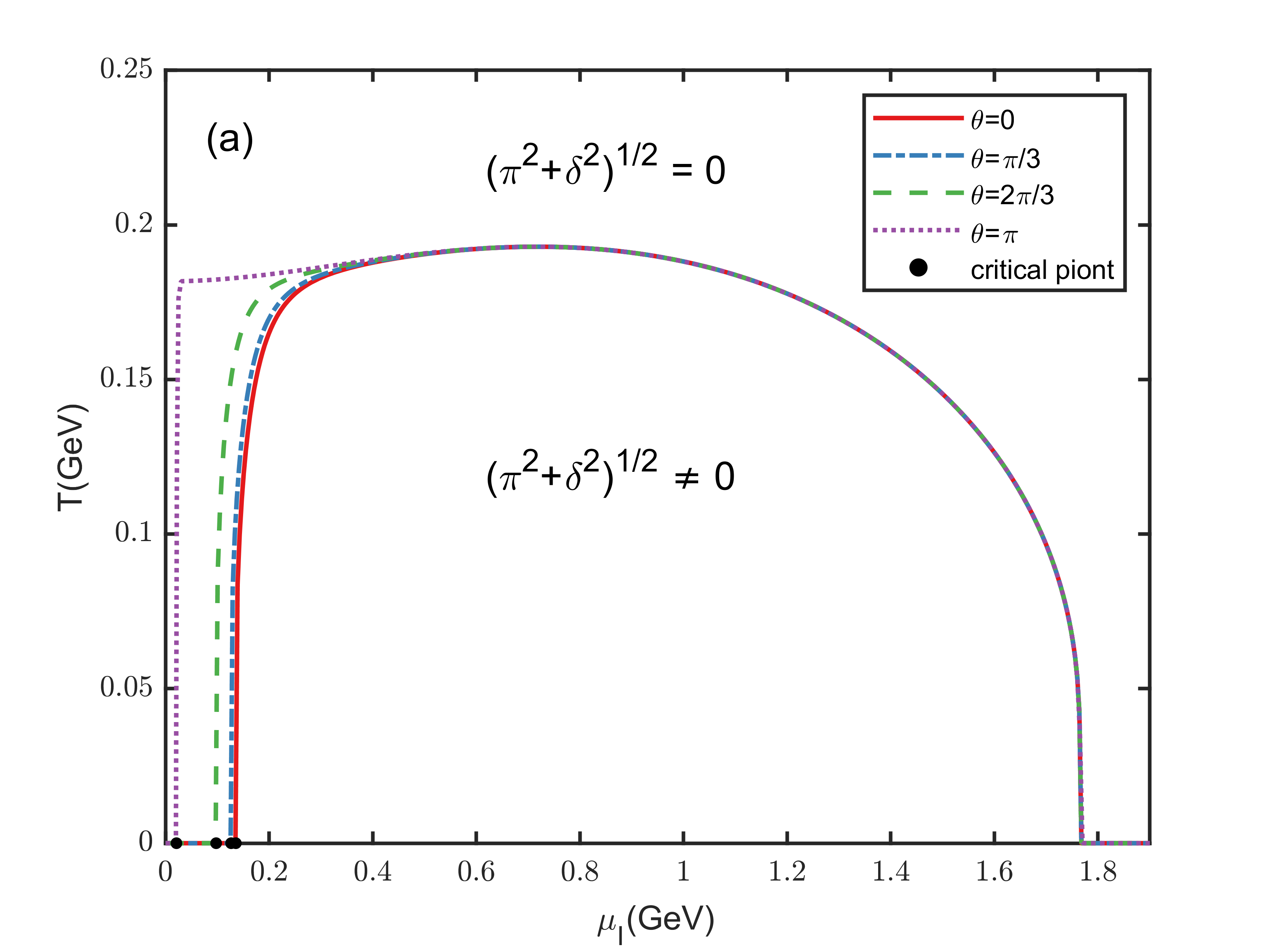}
        \phantomcaption
        \label{fig4a}
    \end{subfigure}
    \hspace{-0.93cm} 
    \begin{subfigure}[b]{0.45\textwidth}
        \includegraphics[width=\textwidth]{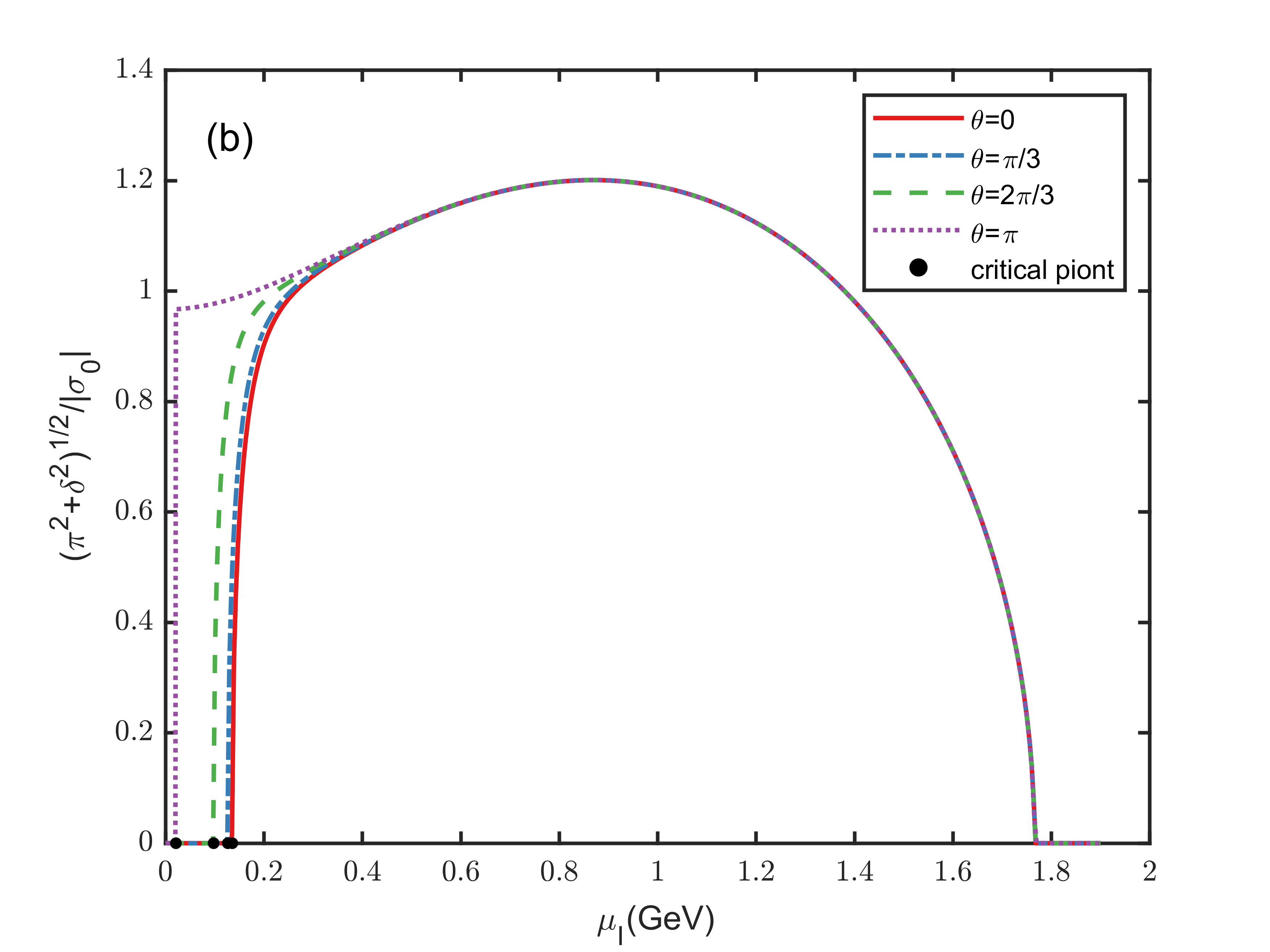}
        \phantomcaption
        \label{fig4b}
    \end{subfigure}
 \caption{The linear relation between the critical temperature $T_c$ and total condensate at zero temperature for several values of $\theta$: (a) The phase diagram in the $T$-$\mu_I$ plane at $\mu_B=0$. (b) The total condensate as function of $\mu_I$ at zero temperature.}
\end{figure*}
Next we present the isospin symmetry breaking phase diagram in the $T$-$\mu_I$ plane at $\mu_B=0$ for different $\theta$ values in Fig.~\ref{fig4a}. It is noteworthy that the criterion for determining the isospin symmetry breaking phase has transitioned from examining the $\pi$ condensate value to evaluating the total condensate $\sqrt{\pi^2+\delta^2}$. We observe that the $\theta$-dependent modulation of the critical isospin chemical potential creates distinct phase boundaries for different $\theta$ values in the $\mu_I < 0.4$ GeV regime. However, for $\mu_I > 0.4$ GeV — where isospin symmetry becomes fully broken for all $\theta$ — the phase boundaries converge completely across different $\theta$ values. 

Analogous to standard BCS theory, the critical temperature $T_c$ and total condensate at zero temperature approximately obey the linear relation:
\begin{equation}
    T_c(\mu_I) \propto \sqrt{\pi^2+\delta^2}(T=0,\mu_I).
\end{equation}
which can be verified by comparing with the $\mu_I$-dependent total condensate profiles under corresponding $\theta$ values shown in Fig.~\ref{fig4b}.

\begin{figure}[htbp]
    \centering
    \includegraphics[width=8.6cm]{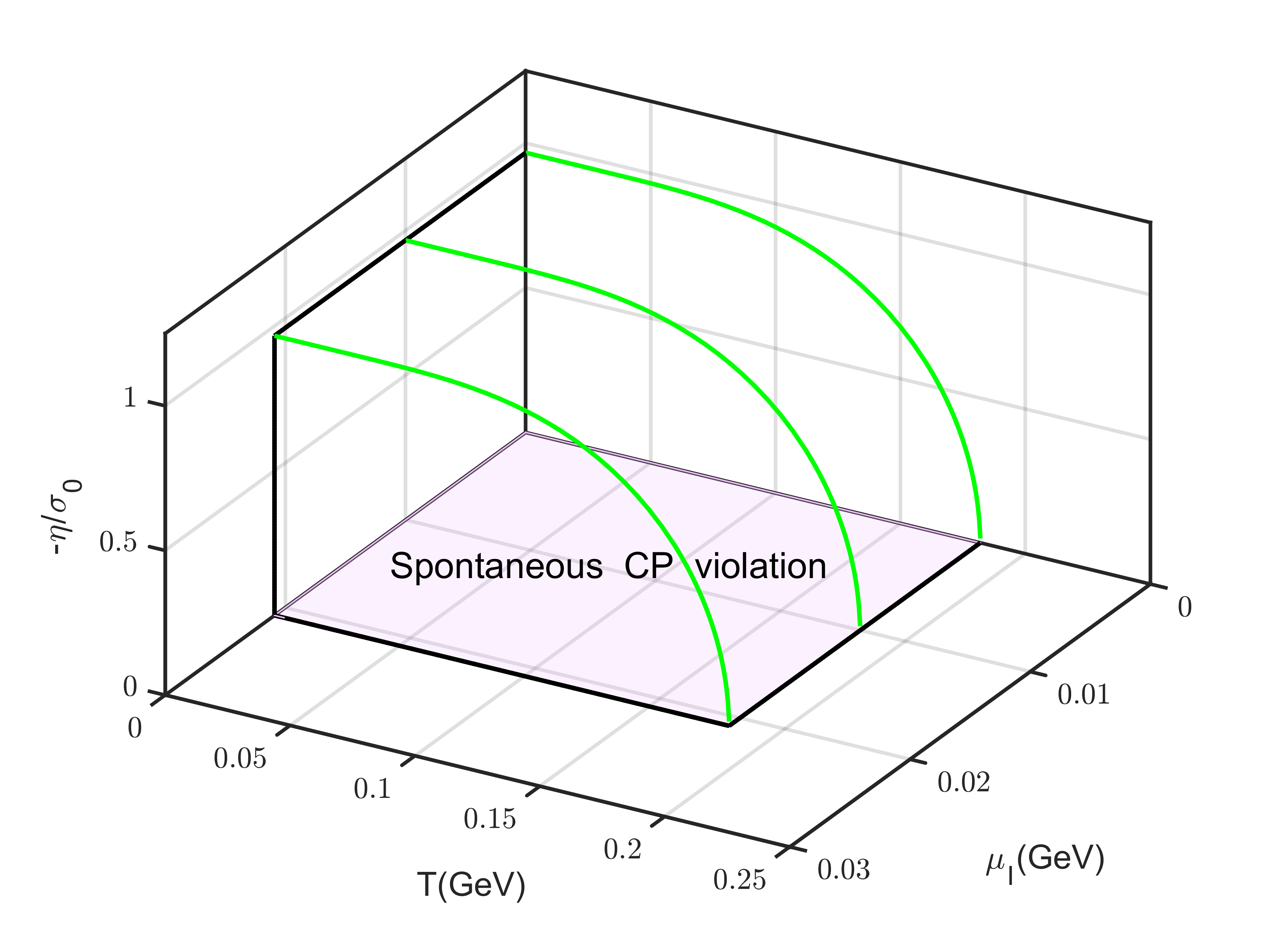}
    \caption{The phase diagram for CP transition in the $T$-$\mu_I$ plane.}
    \label{fig5}
\end{figure}
Finally, we present the CP violation transition phase diagram in the $T$-$\mu_I$ plane in Fig.~\ref{fig5}. As previously discussed, the isospin symmetry breaking phase inherently preserves CP symmetry under pure isospin effects, and this conclusion remains valid at finite temperatures since thermal fluctuations tend to restore symmetries universally. The second-order CP restoration transition lines coincide completely for $\mu_I < 0.021$ GeV, forming a rectangular spontaneous CP violation region bounded by $T=0.182$ GeV and $\mu_I=0.021$ GeV in the phase diagram, and the tricritical point which connects the first and second order phase transitions is located at $(T_c,\mu_I^\text{crit})=(0.182, 0.021)$ GeV. 

\section{finite temperature and baryon density}
~\label{sec:results3}
We now turn to the discussion $\theta$ effects at finite baryon chemical potential $\mu_B$ and temperature $T$ with $\mu_I=0.15$ GeV to keep the system in the initial phase of isospin symmetry breaking as we have investigated it at finite isospin chemical potential $\mu_I$ and temperature $T$.

\begin{figure*}[htbp]
    \centering
    \begin{subfigure}[b]{0.45\textwidth}
        \includegraphics[width=\textwidth]{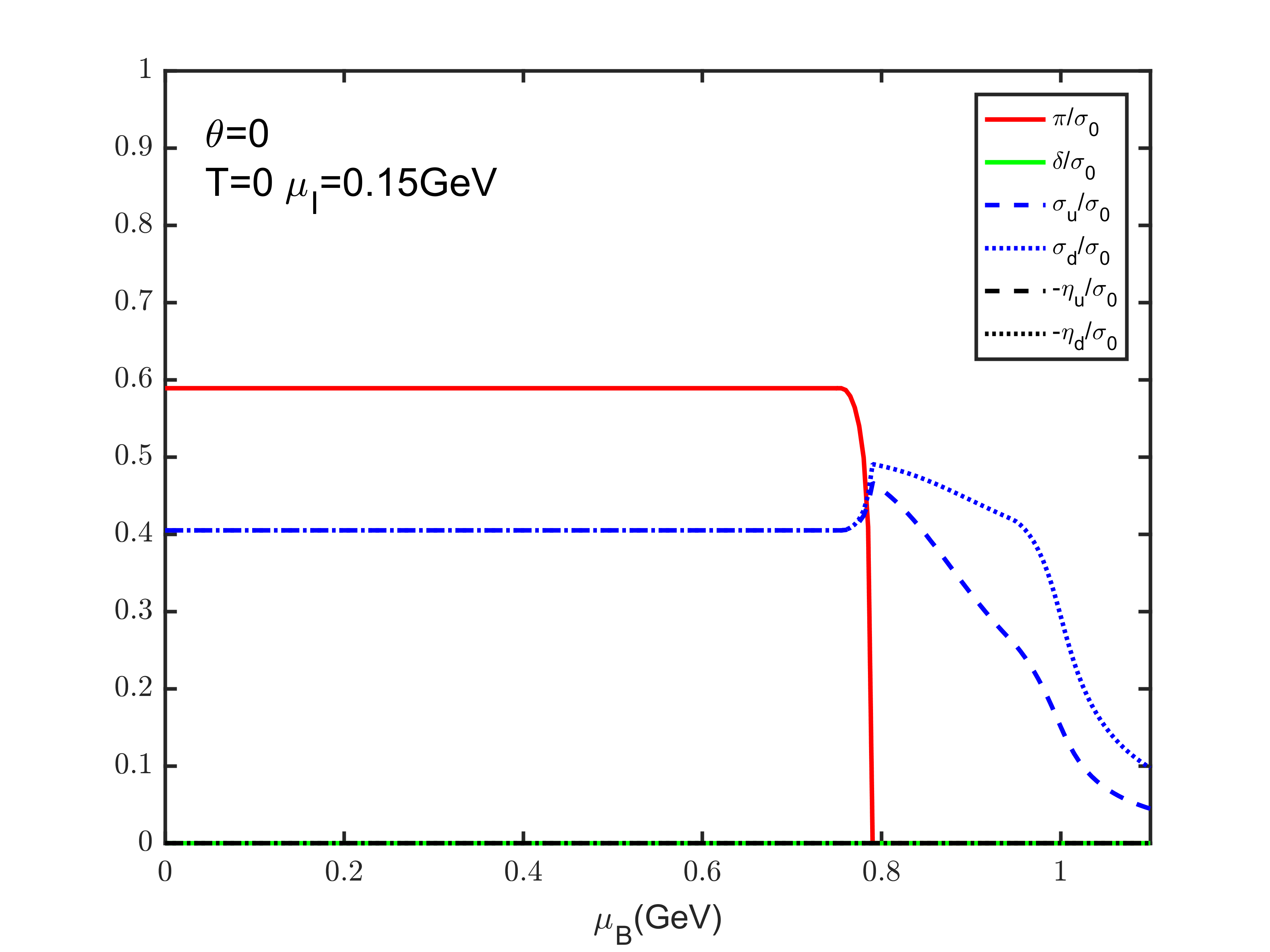}
    \end{subfigure}
    \hspace{-0.93cm} 
    \begin{subfigure}[b]{0.45\textwidth}
        \includegraphics[width=\textwidth]{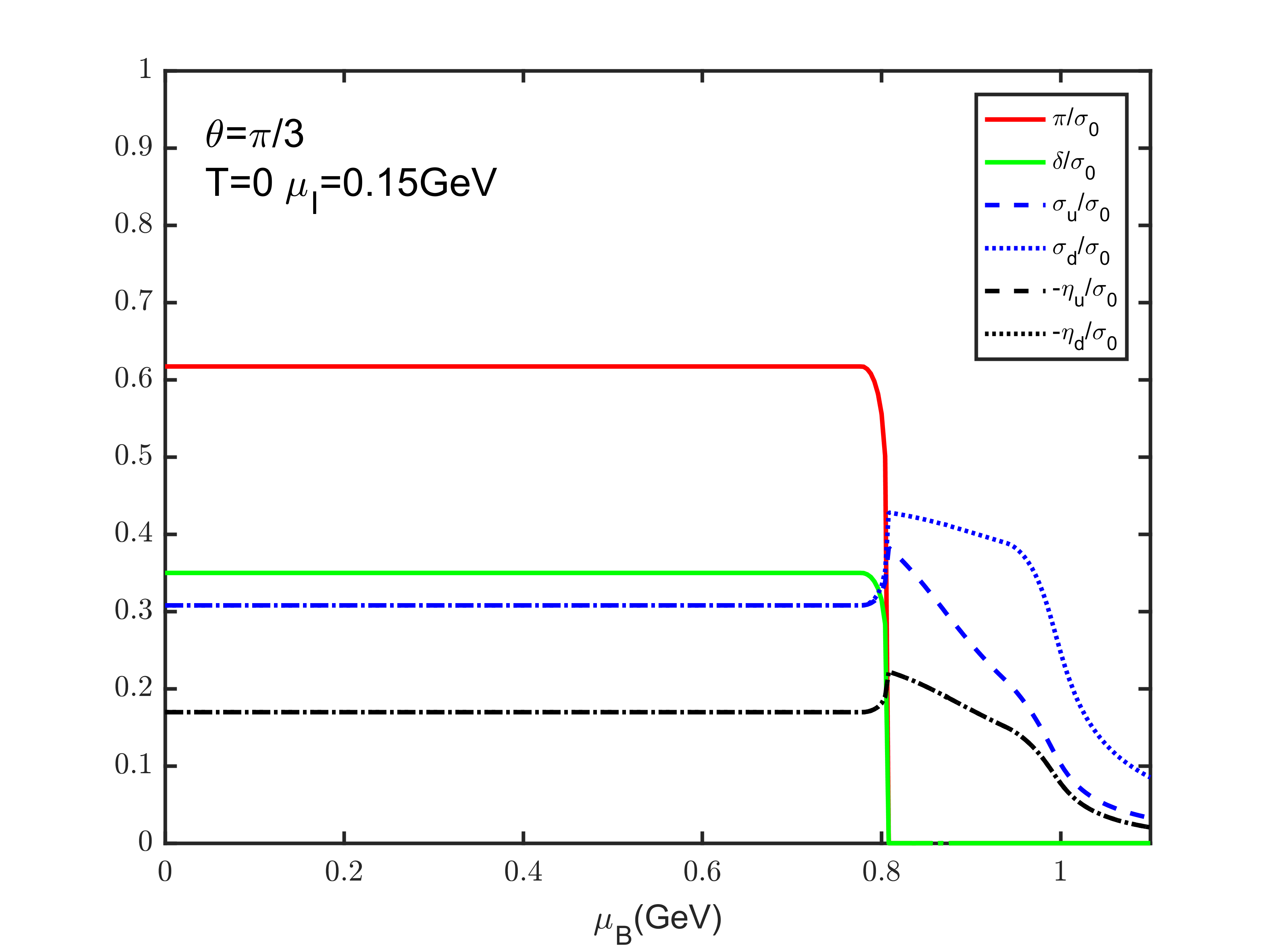}
    \end{subfigure}

    \vspace{0cm} 

    \begin{subfigure}[b]{0.45\textwidth}
        \includegraphics[width=\textwidth]{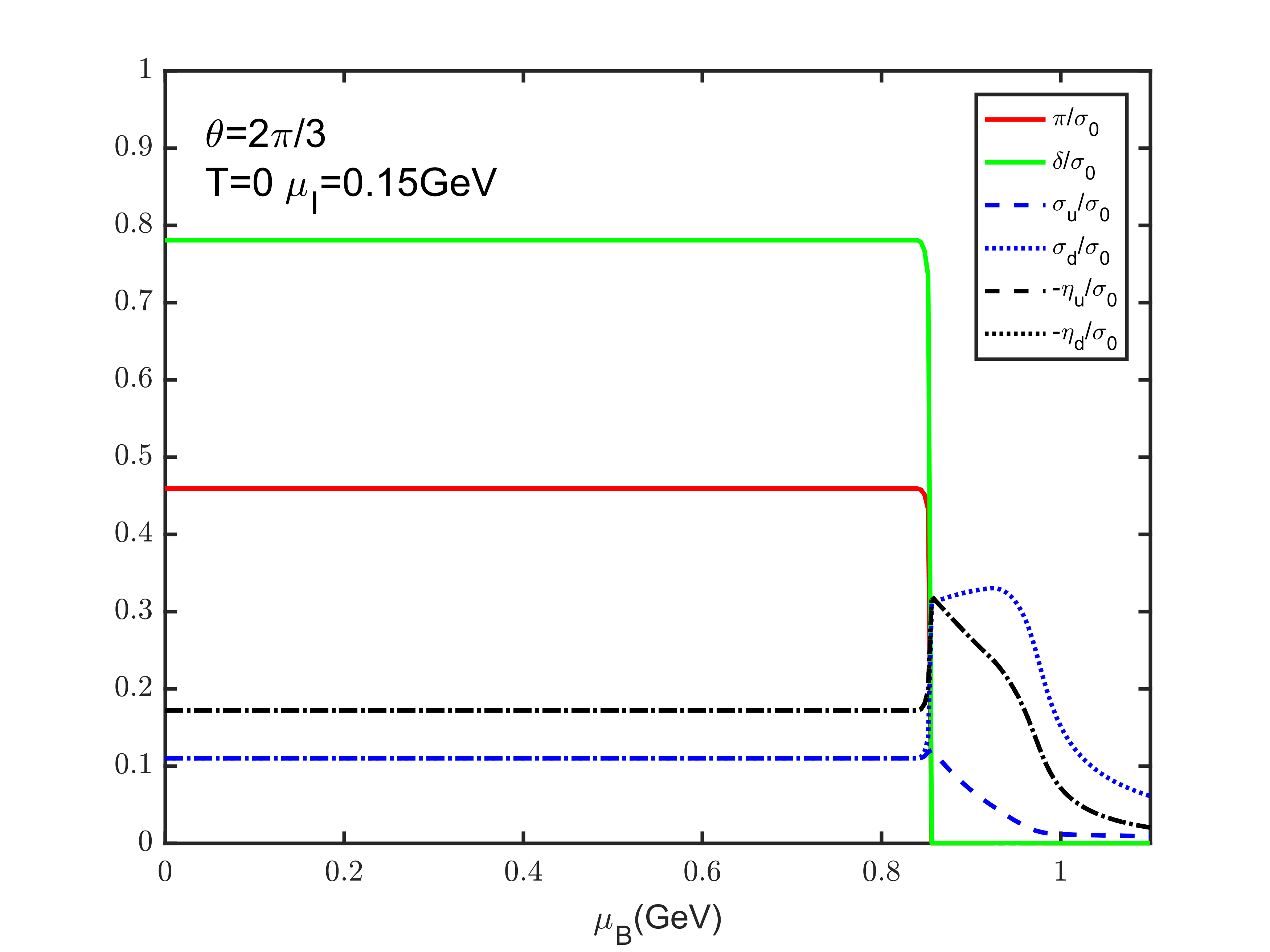}
    \end{subfigure}
   \hspace{-0.93cm} 
    \begin{subfigure}[b]{0.45\textwidth}
        \includegraphics[width=\textwidth]{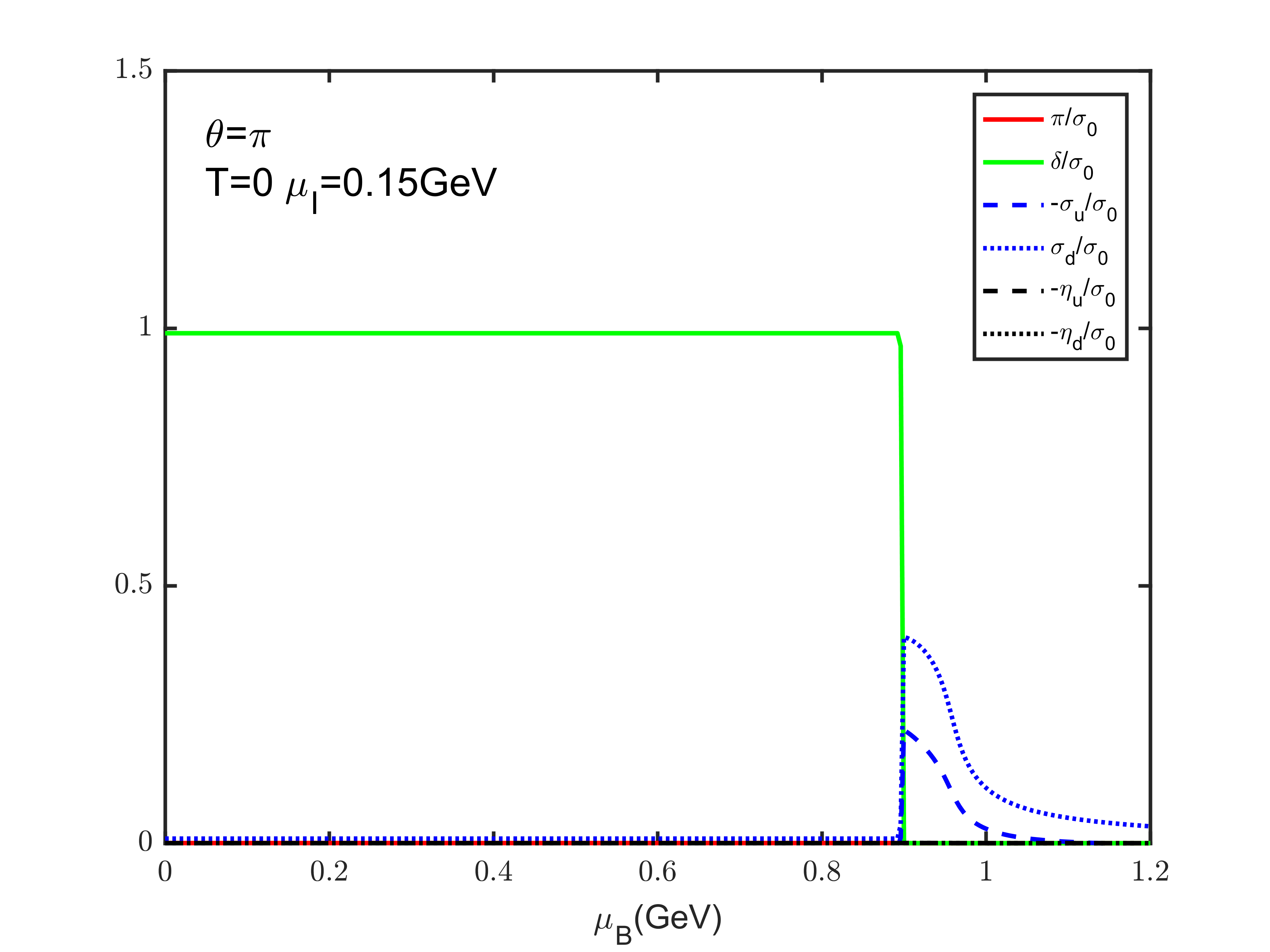}
    \end{subfigure}

    \caption{The four condensates $\sigma$, $\pi$, $\eta$ and $\delta$ scaled by the
chiral condensate $\sigma_0$, as function of baryon chemical potential at $T=0$ and $\mu_I=0.15$ GeV for several values of $\theta \equiv a/f_a$. Top left plot corresponds to $\theta=0$, top right to $\theta=\pi/3$, bottom left to $\theta=2\pi/3$, and finally bottom right to $\theta=\pi$.}
    \label{fig6}
\end{figure*}
The $\mu_B$ dependence of the four condensates at different $\theta$ values, normalized by the vacuum chiral condensate $\sigma_0$ at fixed $T=0$, is shown in Fig.~\ref{fig6}. 

In Fig.~\ref{fig6} ($\theta=0$), only the $\sigma_q$ and $\pi$ condensates are nonzero. Consistent with Ref.~\cite{PhysRevD.71.116001}, the $\sigma_u$ and $\sigma_d$ condensates remain nearly identical, preserving their vacuum values in the pion superfluidity phase. Near the critical $\mu_B^\text{crit}=0.789$ GeV, the $\pi$ condensate begins to decrease while the $\sigma_q$ condensate starts to increase. A first-order phase transition then occurs: the $\pi$ condensate discontinuously drops to zero, while $\sigma_u$ and $\sigma_d$ jump to higher values with $|\sigma_d| > |\sigma_u|$. Subsequently, both chiral condensates decrease with increasing $\mu_B$, exhibiting significant splitting in the normal phase.

In Fig.~\ref{fig6} ($\theta=\pi/3$), as $\theta$ increases, the $\eta_q$ and $\delta$ condensates develop nonzero values. All six condensates maintain their vacuum values in the isospin symmetry-breaking phase. At a higher critical baryon chemical potential $\mu_B^\text{crit}=0.808$ GeV, the $\pi$ and $\delta$ condensates drop discontinuously to zero, while the $\sigma_q$ and $\eta_q$ condensates jump to larger values. This elevated critical $\mu_B$ arises from $\theta$-enhanced spontaneous isospin symmetry breaking. Subsequently, the chiral condensates $\sigma_u$ and $\sigma_d$ decrease with increasing $\mu_B$, exhibiting more significant splitting, while the $\eta_q$ condensates remain nearly identical across the calculated range.

In Fig.~\ref{fig6} ($\theta=2\pi/3$), the condensates exhibit analogous behavior with one key distinction: $\sigma_u$ first increases then decreases in the $\sqrt{\pi^2+\delta^2}=0$ phase with a critical baryon chemical potential $\mu_B^\text{crit}=0.856$ GeV. 

In Fig.~\ref{fig6} ($\theta=\pi$), the $\delta$ condensate dominates absolutely, while the scalar condensate $\sigma$ retains small nonzero values. Both $\eta$ and $\pi$ condensates remain fully suppressed as in previous cases. The $\delta$ condensate drops to zero at critical $\mu_B^\text{crit}=0.899$ GeV, with only $\sigma_q$ discontinuously emerging. Crucially, $\sigma_u$ evolves oppositely to $\sigma_d$, a phenomenon likely driven by $\theta$-amplified isospin asymmetry effects and explicit
chiral symmetry breaking.

\begin{figure*}[htbp]
    \centering
    \begin{subfigure}[b]{0.45\textwidth}
        \includegraphics[width=\textwidth]{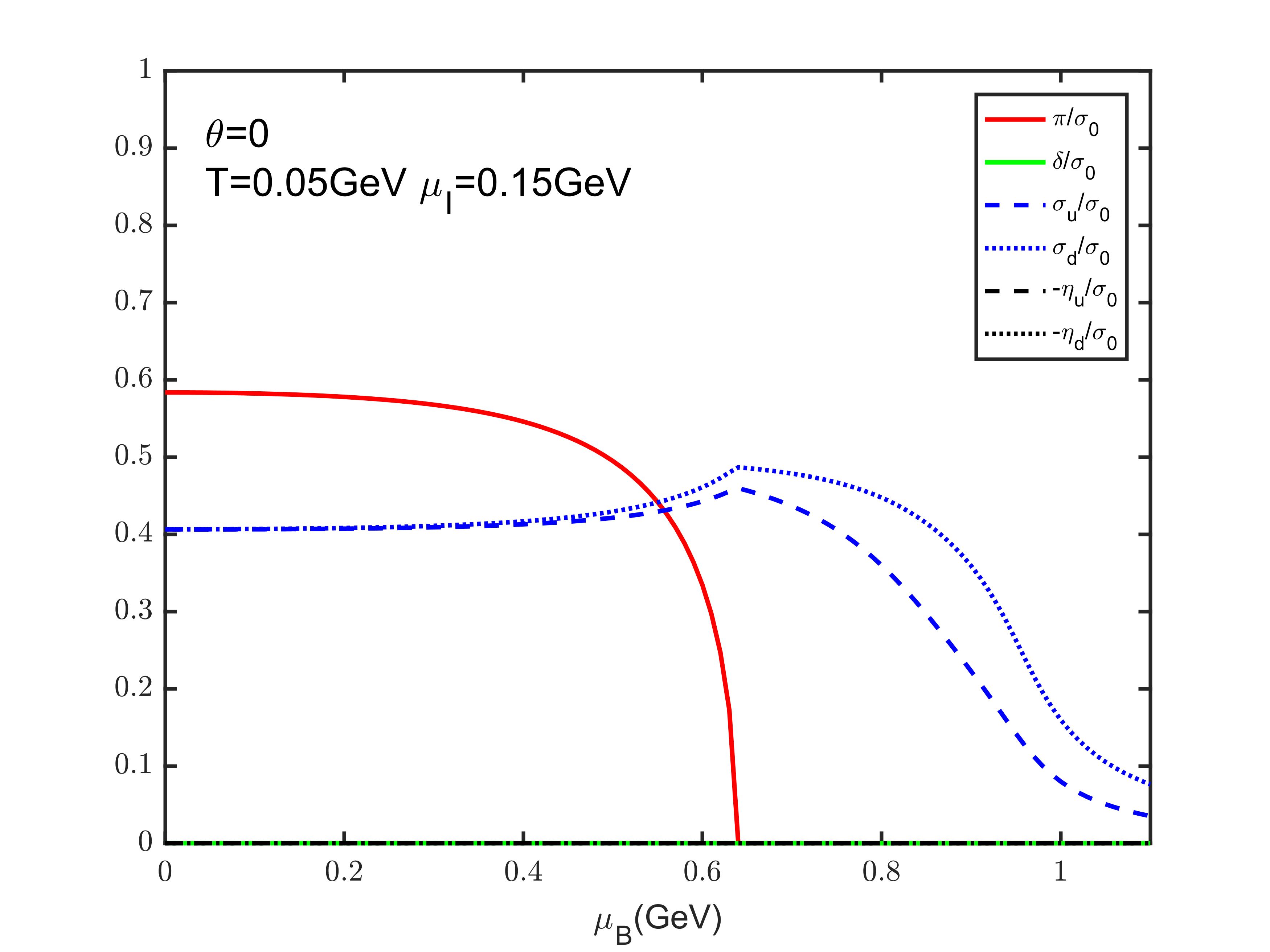}
    \end{subfigure}
    \hspace{-0.93cm} 
    \begin{subfigure}[b]{0.45\textwidth}
        \includegraphics[width=\textwidth]{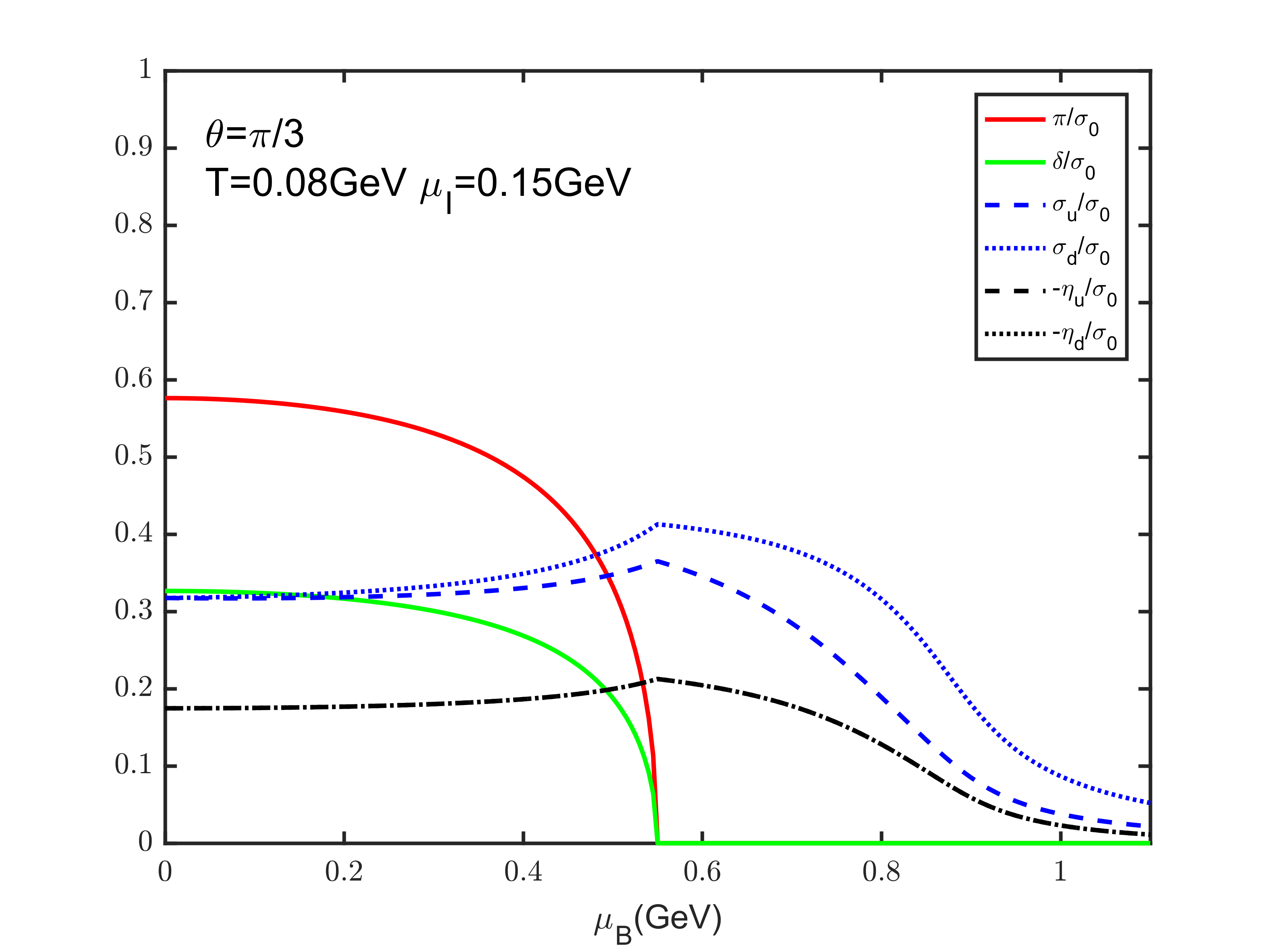}
    \end{subfigure}

    \vspace{0cm} 

    \begin{subfigure}[b]{0.45\textwidth}
        \includegraphics[width=\textwidth]{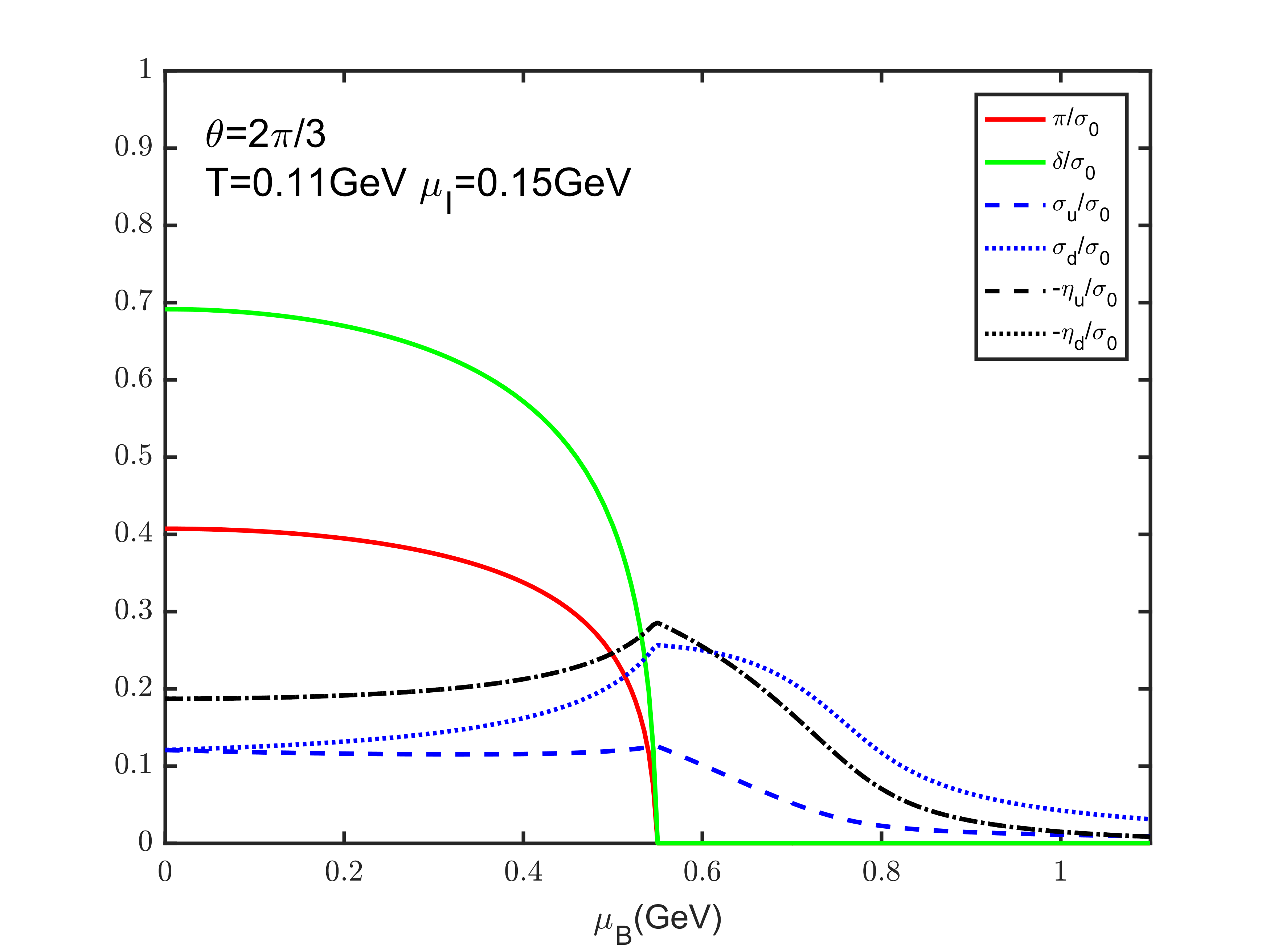}
    \end{subfigure}
    \hspace{-0.93cm} 
    \begin{subfigure}[b]{0.45\textwidth}
        \includegraphics[width=\textwidth]{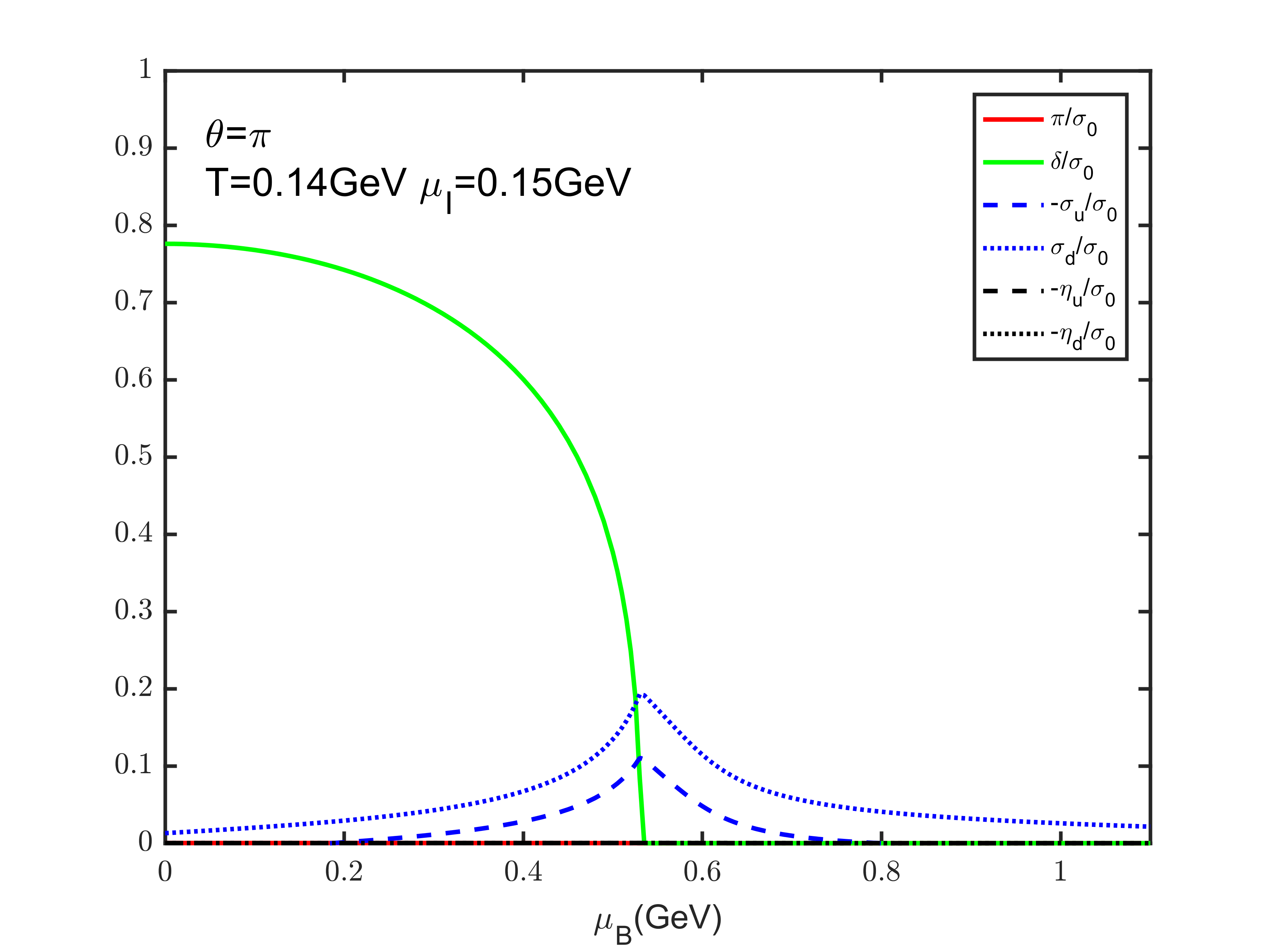}
    \end{subfigure}

    \caption{The four condensates $\sigma$, $\pi$, $\eta$ and $\delta$ scaled by the
chiral condensate $\sigma_0$, as function of baryon chemical potential at finite $T$ and fixed $\mu_I=0.15$ GeV for several values of $\theta \equiv a/f_a$. Top left plot corresponds to $\theta=0$, top right to $\theta=\pi/3$, bottom left to $\theta=2\pi/3$, and finally bottom right to $\theta=\pi$.}
    \label{fig7}
\end{figure*}
When extending to finite temperatures in Fig.~\ref{fig7}, we observe that the temperature effect converts the phase transition from first-order to second-order when $T$ exceeds a threshold value, concurrently reducing the critical baryon chemical potential. Additionally, the temperature effect induce significant splitting between the two chiral condensates ($\sigma_u$, $\sigma_d$) not only in the normal phase but also in the $\sqrt{\pi^2+\delta^2} \neq 0$ regime, while the $\eta_q$ condensates remain identical throughout.

\begin{figure}[htbp]
    \centering
    \includegraphics[width=8.6cm]{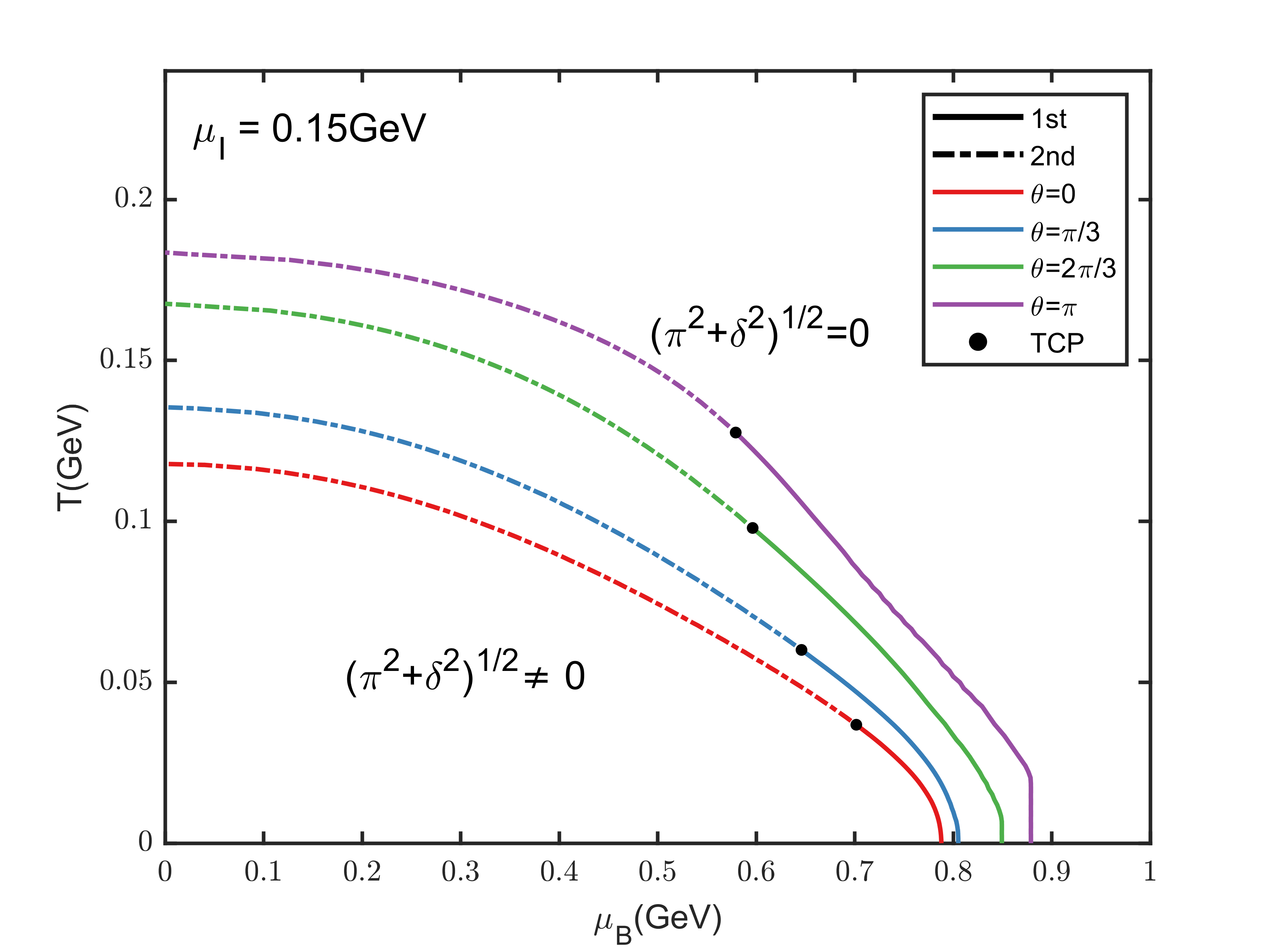}
    \caption{The phase diagram of isospin symmetry breaking in the $T$-$\mu_B$ plane at $\mu_I=0.15$ GeV for several values of $\theta$. The solid part of the line indicates a first order phase transition while the dashed part indicates one of
second order. The black dots are the positions of the tricritical point.}
    \label{fig8}
\end{figure}
The $T$-$\mu_B$ phase diagram at fixed $\mu_I$ for different $\theta$ values is shown in Fig.~\ref{fig8}. Second-order transitions dominate the high-temperature regime, whereas first-order transitions persist at high $\mu_B$. Increasing $\theta$ expands the phase boundary outward due to amplified isospin asymmetry effects, maintaining constant first-order critical $\mu_B^\text{crit}$ values across an extended low-temperature region. As tabulated in Table~\ref{tab1}, the tricritical point connecting first and second-order transitions shifts toward higher $T$ and lower $\mu_B$ with increasing $\theta$.
\begin{table*}[htbp]
\centering
\caption{Values of the tricritical points at various $\theta$.  }
\label{tab1}
\renewcommand\arraystretch{1.5}
\setlength{\tabcolsep}{5mm}
\begin{tabular}{ccccc}
\hline\hline
$\theta$ & $0$ & $\frac{\pi}{3}$ & $\frac{2\pi}{3}$ &$\pi$\\
\hline
$(T_c, \mu_B^{\text{crit}})$ & (0.037, 0.702) GeV &(0.060, 0.646) GeV &(0.098, 0.596) GeV&(0.128, 0.512) GeV \\
\hline\hline
\end{tabular}
\end{table*}

\section{Axion effects on Bulk properties and the structure of nonstrange quark stars}
~\label{sec:results4}
To investigate the $\theta$ effects on bulk properties of two flavor quark matter, we can study the axion effects on the massive compact stars which consists of nonstrange quark matter with the assumption that two flavor quark matter is stable.

Including diquark pairing contributions, the computed mass-radius relationship for compact stars containing color superconducting quark matter \cite{Alford1998,Rapp1998} demonstrates convergence with results from unpaired quark matter models. This agreement stems from the energy scale of the diquark condensate being orders of magnitude below the corresponding Fermi energy \cite{Shovkovy2003,Stefan2004,AlfordReddy2003}. Consequently, we explicitly exclude diquark pairing interactions to maintain focus on primary order parameters. 

To ensure computational validity in this study, the momentum cutoff $\Lambda$ must be set to satisfy $\Lambda > \mu_B/3$, where the baryon chemical potential in massive quark star interiors does not exceed $\mu_B \lesssim 1.8\,\text{GeV}$ based on our model calculations. 

Considering the electrical neutrality of  quark stars and electroweak reactions in them, we should take the $\beta$ equilibrium and electric charge neutrality conditions into account, the conditions are
\begin{gather}
    \mu_d=\mu_u+\mu_e,\\
    \frac{2}{3}\rho_u-\frac{1}{3}\rho_d-\rho_e=0.
\end{gather}
Particle number densities depend on their chemical potentials, therefore there is only one independent variable in these chemical potentials. As we have obtained the effective thermodynamic potential at finite chemical potential $\Omega(a,\mu_B,\mu_I)$ with the physical vacuum $\Omega(a,0,0)$, the pressure relative to the physical vacuum, and energy density are given by
\begin{gather}
    P=\Omega(a,0,0)-\Omega(a,\mu_B,\mu_I)+\frac{\mu_e^4}{12\pi^2},\label{P}\\
    \epsilon=-p+\mu_B\rho_B+\mu_I\rho_I+\frac{\mu_e^4}{3\pi^2}.
\end{gather}

It is noticed that in Eq.(\ref{P}), the zero-point pressure subtraction is at finite $\theta$ not at $\theta=0$. As a candidate of dark matter, axions are weakly coupled to quark matter. Under gravitational influence, the axion is distributed both inside and outside the compact star, thus making the compact star immersed in a medium of axion background. The difference pressure at finite $\theta$ and at $\theta=0$ is similar to a bag constant.

\begin{figure}[htbp]
    \centering
    \includegraphics[width=8.6cm]{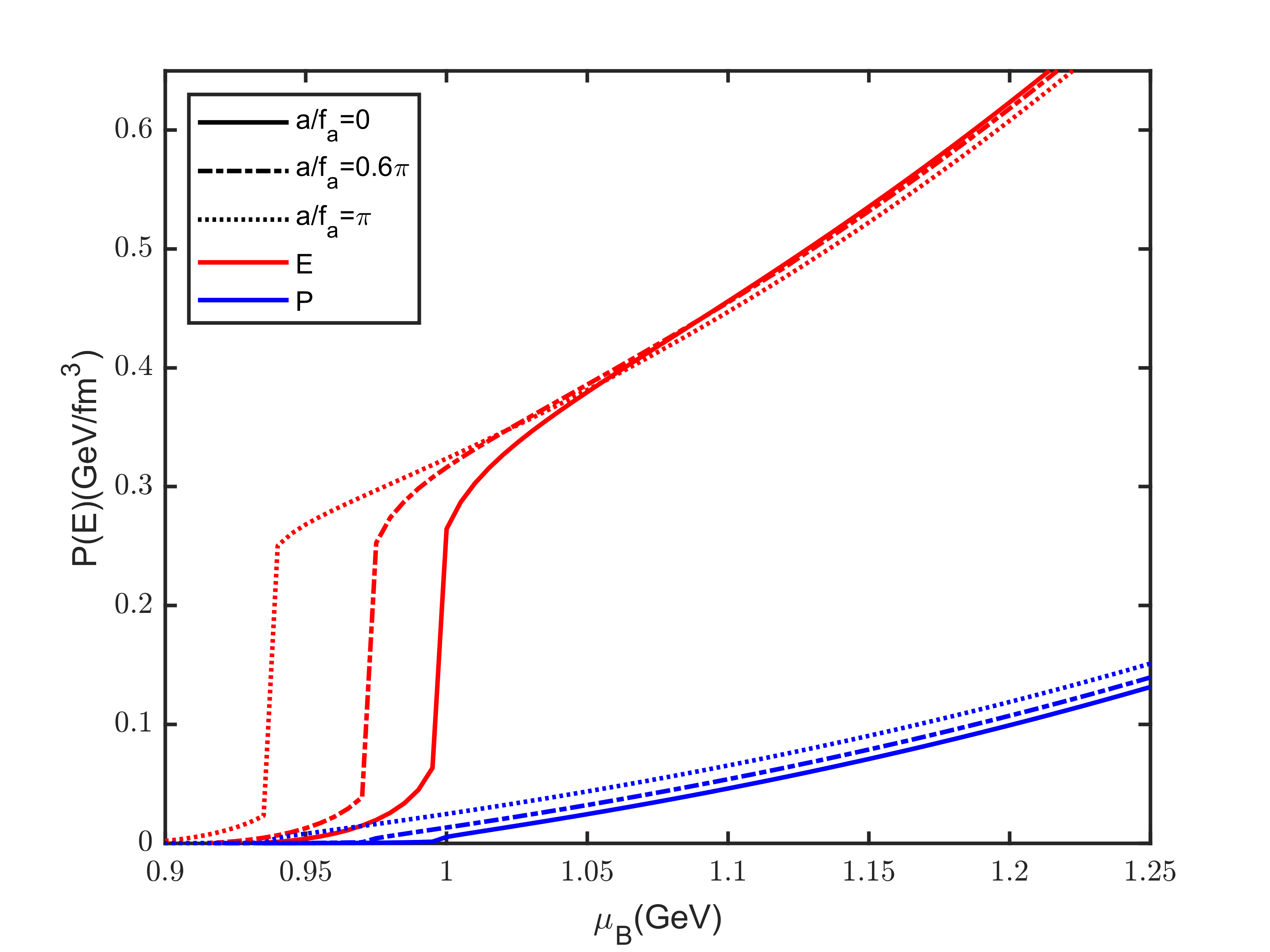}
    \caption{The pressure and energy density varying with $\mu_B$ in dense QCD system with conditions of $\beta$ equilibrium and electric neutrality, computed for several values of the scaled axion field $a/f_a$.}
    \label{fig9}
\end{figure}
By solving the $\beta$ equilibrium and electric charge neutrality conditions, the pressure and energy density as functions of $\mu_B$ are displayed in Fig.~\ref{fig9}. It can be seen that as the scaled axion field $a/f_a$ increases, the first order phase transition in quark matter moves to the lower baryon chemical potential, resulting in the earlier appearance of chiral phase transition and free quarks so that the EOS of quark matter becomes stiffer. It should be noted that, under these parameters, there exists an interval with both energy density and pressure prior to the occurrence of the first-order phase transition. This will lead to a mass-radius relationship similar to that of hybrid stars \cite{XU2021115540}. However, since the thermodynamic potential within the mean-field approximation does not incorporate degrees of freedom from hadrons, it is still referred to as a quark star in this paper. 
In the high-density quark-gluon plasma (QGP) phase, where $\theta$ couples to condensates (e.g., chiral and pseudoscalar $\eta$ condensates), all relevant condensates approach zero. Consequently, the thermodynamic potentials for different $\theta$ values become essentially identical in this regime. The pressure difference between distinct $\theta$ configurations in this phase is therefore determined solely by the vacuum thermodynamic potential difference:
\begin{equation}
\Delta P \approx \Omega_{\text{vac}}(\theta_1) - \Omega_{\text{vac}}(\theta_2).
\end{equation}

As generally done, we assume the nonstrange star to be a spherically symmetric object. The mass-radius relation of hybrid stars can be obtained by solving the general relativistic equation of hydrostatic equilibrium known as the Tolman-Oppenheimer-Volkoff (TOV) equation
\begin{gather}
    \frac{dP(r)}{dr} = -\frac{G(\epsilon + P)(M + 4\pi r^3 P)}{r(r - 2GM)},\\
    M(r)= \int_0^r  dr'4\pi r'^2\epsilon(r').
\end{gather}

\begin{figure}[htbp]
    \centering
    \includegraphics[width=8.6cm]{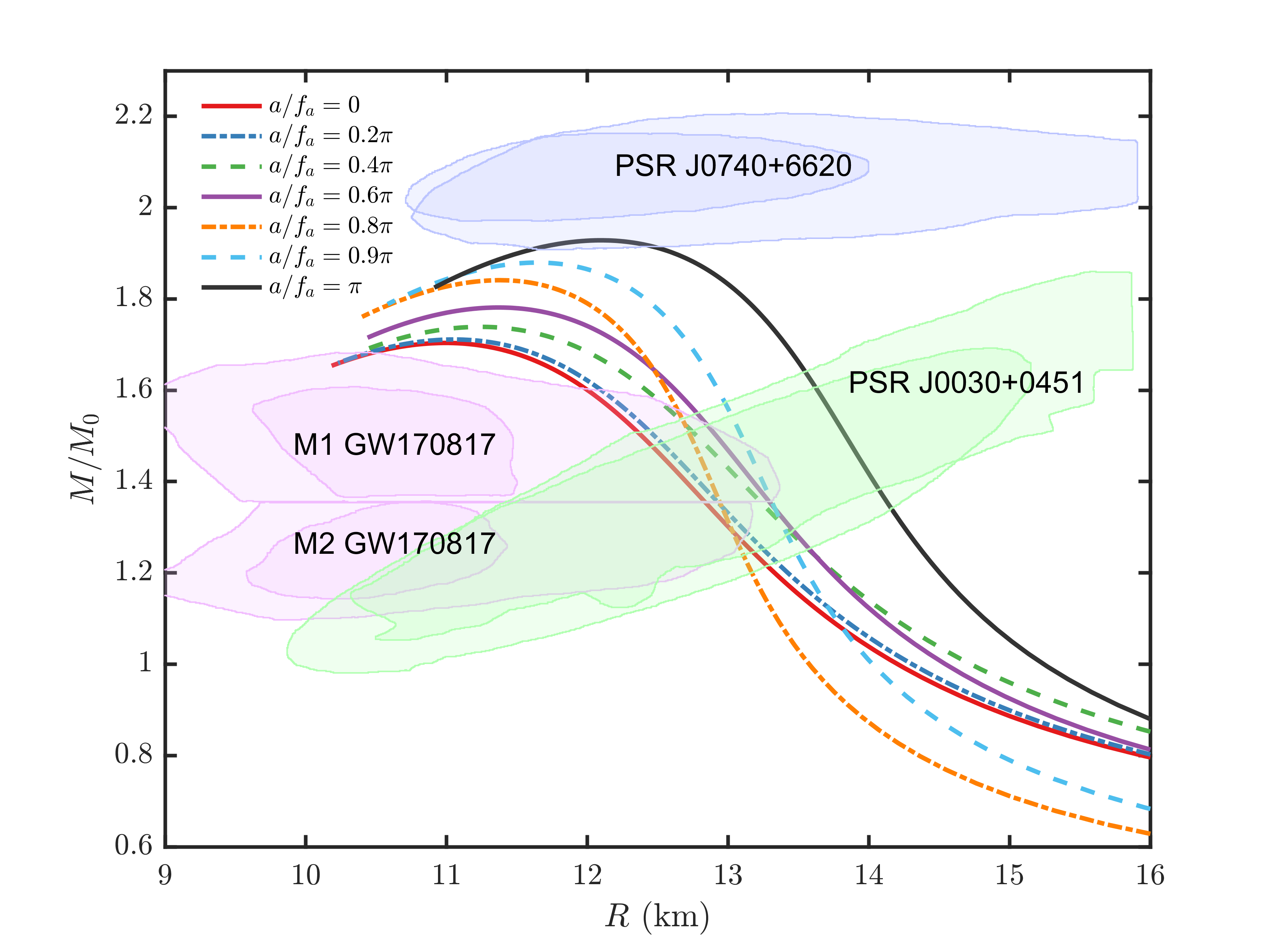}
    \caption{The mass and radius relation of nonstrange quark stars, computed for several values of the scaled axion field $a/f_a$. The constraints from multimessenger astronomy observations are shown by shaded regions.}
    \label{fig10}
\end{figure}

The mass-radius relation for nonstrange quark stars incorporating the scaled axion field $a/f_a$ is presented in Fig.~\ref{fig10}. Observational constraints derived from Bayesian analysis of pulse profile data for PSR~J0030+0451~\cite{Riley2019,Miller2019} and PSR~J0740+6620~\cite{Riley2021,Miller2021}, along with the analyses of the gravitational wave signal from the NS merger GW170817~\cite{Abbott2017} are superimposed for comparison. Key findings include:
\begin{itemize}
    \item A systematic increase in maximum mass ($M_{\rm max}$) and corresponding radius ($R_{M_{\rm max}}$) with axion coupling strength $a/f_a$, consistent with equation-of-state stiffening effects. But its effects on hybrid star matter demonstrate counterintuitive evolutionary behavior in the EOS of  quark matter for a decrease in both the speed of sound and the polytropic index during the hadron-quark transition~\cite{PhysRevD.111.L051501}.
    \item Compatibility of all parameter sets with multi-messenger constraints from PSR~J0030+0451, GW170817, and PSR~J0740+6620 (for $a/f_a\sim\pi$).
\end{itemize}
In addition, when considering the rotation of the quark star, the value of the maximum mass is roughly $10\%–20\%$ higher than that in the non-rotating case \cite{Li2016}.

\section{Summary and outlook}
\label{sec:summary}
It is well established that a nonzero CP-violating $\theta$ parameter induces richer phase structures in Quantum Chromodynamics, with the $\theta=\pi$ case exhibiting spontaneous CP violation through the Dashen phenomenon. This motivates detailed studies of $\theta$-modified phase transitions in hot and dense quark matter. Building upon existing research on $\theta$-dependent chiral symmetry breaking, deconfinement, and color superconductivity, our work extends the understanding of theta vacuum physics to systems with finite isospin chemical potential ($\mu_I$), mapping CP symmetry restoration across the $T$-$\mu_I$ phase diagram. Within the two-flavor Nambu-Jona-Lasinio framework, we first review the physical origin of the Kobayashi-Maskawa-'t Hooft determinant interaction, which emulates the topological Chern-Simons term in low-energy QCD. Notably, the bubble summation structure in the random phase approximation (RPA) becomes significantly more complex in the isospin symmetry-broken phase compared to the chiral-broken regime. By deriving analytical expressions for the thermodynamic potential $\Omega$ and quark propagators $S$, we circumvent numerical matrix inversion and determinant calculations in solving gap equations. The condensates both in the scalar as well as the pseudoscalar channel are considered. 

In systems with only isospin chemical potential ($\mu_I$), the introduction of the $\theta$-term induces the following effects: Firstly, the $\theta$ parameter significantly enhances the magnitude of ($\eta$, $\delta$) meson condensates while suppressing the strength of ($\sigma$, $\pi$) meson condensates. Secondly, the $\theta$ effect promotes spontaneous isospin symmetry breaking by reducing the critical threshold $\mu_I^{\text{crit}}$ for spontaneous isospin symmetry breaking. When $\theta=\pi$, the critical isospin chemical potential decreases to $\mu_I^{\text{crit}} = 0.021\ \mathrm{GeV}$, where the system undergoes a first-order phase transition. Notably, during this process, the ($\sigma$, $\pi$) condensates and ($\eta$, $\delta$) condensates exhibit analogous evolutionary behavior except $\theta=0$ or $\pi$.

Extending the investigation to systems of finite temperature ($T$) and finite baryon chemical potential ($\mu_B$) reveals that these $\theta$-induced effects persist. Due to the $\theta$-enhanced spontaneous isospin symmetry breaking, the phase boundaries of the isospin-broken phase expand in both $T$-$\mu_I$ and $T$-$\mu_B$ phase diagrams. Specifically, for the spontaneous CP symmetry-breaking phase transition occurring at $\theta=\pi$, we successfully extend its existence region from the $T$-$\mu_B$ phase diagram to the $T$-$\mu_I$ phase diagram.

In studying axion field effects on nonstrange quark star matter, we find that the baryon chemical potential $\mu_B$ corresponding to the first-order chiral phase transition decreases with increasing $a / f_a$. This leads to earlier emergence of free quarks and enhances the stiffness of the quark matter equation of state. While the axion's influence on pure nonstrange quark star matter aligns with theoretical expectations, its effects on hybrid star matter (containing coexisting quark-nucleon phases) demonstrate counterintuitive evolutionary behavior in the equation of state.

Last but not least, we may estimate the axion mass and self-coupling at finite isospin chemical potential in the near future.

\begin{acknowledgments}
We acknowledge helpful discussions with Xu-Guang Huang, Jun Xu, Zhen-Yan Lu and JingDong Shao.
This work is supported in part by the National Natural Science Foundation of China (NSFC) Grant Nos: 12235016, 12221005, and the China Postdoctoral Science Foundation under Grant No. 2024M750489.
\end{acknowledgments}

\appendix
\section{THE LAGRANGIAN FROM MEAN FIELD APPROXIMATION}
\label{A}
In the mean-field approximation, it is assumed that the deviations due to fluctuations of all quantities $F$ from their thermal average values $\langle F\rangle$ are small. Thus, the following relations can be introduced to linearize the Lagrangian
\begin{align}
    (\bar\psi \Gamma\psi) &\approx \langle \bar\psi \Gamma\psi \rangle, \label{a1}\\
(\bar\psi \Gamma\psi)^2 &\approx 2 \langle \bar\psi \Gamma\psi \rangle (\bar\psi \Gamma\psi ) - \langle \bar\psi \Gamma\psi \rangle^2 ,\\
(\bar\psi \Gamma_i\psi)(\bar\psi \Gamma_j\psi) &\approx ( \bar\psi \Gamma_i\psi ) \langle \bar\psi \Gamma_j\psi \rangle + \langle \bar\psi \Gamma_i\psi \rangle ( \bar\psi \Gamma_j\psi ) \notag\\
&- \langle \bar\psi \Gamma_i\psi \rangle \langle \bar\psi \Gamma_j\psi \rangle. \label{a2}
\end{align}
In the mean-field approximation by using the relations of Eqs. (\ref{a1})-(\ref{a2}), the following term can be expressed as
\begin{gather}
    \sum_{a=1}^{2} ( \bar{\psi} i \gamma_5 \tau^a \psi )^2= 2\bar{\psi}  i \gamma_5 \tau_1 \pi \psi - \pi^2, \\
     \sum_{a=1,2}(\bar{\psi}\tau^a \psi)^2=2\bar{\psi} \tau_1 \delta \psi - \delta^2,\\
    \sum_{a=0,3} ( \bar{\psi} i \gamma_5 \tau^a \psi )^2=4\bar{\psi}i\gamma_5 diag(\eta_u,\eta_d)\psi-2(\eta_u^2+\eta_d^2),\\
    \sum_{a=0,3}(\bar{\psi}\tau^a \psi)^2=4\bar{\psi} diag(\sigma_u,\sigma_d)\psi-2(\sigma_u^2+\sigma_d^2),\\
    \det(\bar{\psi}\Gamma \psi)=\bar{\psi} \Gamma \begin{pmatrix}
        \langle\bar{d}\Gamma d\rangle &-\langle\bar{d}\Gamma u\rangle \\
        -\langle\bar{u}\Gamma d \rangle &\langle\bar{u}\Gamma u \rangle
    \end{pmatrix} \psi\notag \\
    \quad\quad\quad\quad-\left(\langle \bar{u}\Gamma u\rangle\langle \bar{d}\Gamma d\rangle 
    -\langle \bar{u}\Gamma d\rangle\langle \bar{d}\Gamma u\rangle\right).
\end{gather}
Then the fermionic interaction term can be expressed as
\begin{gather}
    \mathcal{L}_{\bar{q}q}=\bar{\psi}\Sigma_s \psi-\nu_s,\\
    \Sigma_s=G \begin{pmatrix}
        4i\gamma^5\eta_u+4\sigma_u&2i\gamma^5\pi+2\delta\\
        2i\gamma^5\pi+2\delta&4i\gamma^5\eta_d+4\sigma_d
    \end{pmatrix},\\
    \nu_s=G\left[ 2\left(\eta_u^2+\eta_d^2\right)+2\left(\sigma_u^2+\sigma_d^2\right)+\pi^2+\delta^2\right],
\end{gather}
where $\nu_s$ is the contribution of the condensate energy from the fermionic interaction and $\Sigma_s$ is the self-energy contributed by the fermionic interaction.

The KMT term in the mean-field approximation with the $\theta$ has some difficulties; to clarify the origin of this term, we proceed as follows
\begin{align}
    \mathcal{L}_{\text{det}}=&-K\left[e^{-i\frac{a}{f_a}}\det\bar{\psi}(1+\gamma^5)\psi+e^{i\frac{a}{f_a}}\det\bar{\psi}(1-\gamma^5)\psi\right]\notag\\  
    =&-K\cdot2Re\left[e^{-i\frac{a}{f_a}}\det\bar{\psi}(1+\gamma^5)\psi\right]\notag\\
    =&-2K\Big\{ \cos(\frac{a}{f_a})Re\left[\det\bar{\psi}(1+\gamma^5)\psi\right]\notag\\
    &\quad\quad\quad+\sin(\frac{a}{f_a})Im\left[\det\bar{\psi}(1+\gamma^5)\psi\right]\Big\}.
\end{align}
Then, we can safely obtain the expression
\begin{widetext}
\begin{gather}
    \mathcal{L}_\text{det}=\bar{\psi}\Sigma_k\psi-\nu_k,\\
     \Sigma_k=-2K\begin{pmatrix}
        \cos(\frac{a}{f_a})(\sigma_d-i\eta_d\gamma^5)-\sin(\frac{a}{f_a})(\eta_d+i\sigma_d\gamma^5)&\left[\cos(\frac{a}{f_a})\frac{\pi}{2} +\sin(\frac{a}{f_a})\frac{\delta}{2} \right]i\gamma^5+\left[\sin(\frac{a}{f_a})\frac{\pi}{2}-\cos(\frac{a}{f_a})\frac{\delta}{2}\right]\\
        \left[\cos(\frac{a}{f_a})\frac{\pi}{2} +\sin(\frac{a}{f_a})\frac{\delta}{2} \right]i\gamma^5+\left[\sin(\frac{a}{f_a})\frac{\pi}{2}-\cos(\frac{a}{f_a})\frac{\delta}{2}\right]&\cos(\frac{a}{f_a})(\sigma_u-i\eta_u\gamma^5)-\sin(\frac{a}{f_a})(\eta_u+i\sigma_u\gamma^5)
        \end{pmatrix},\\
    \nu_k=-2K\left[\cos(\frac{a}{f_a})\left(\sigma_u\sigma_d-\eta_u\eta_d-\frac{\delta^2}{4}+\frac{\pi^2}{4}\right)-\sin(\frac{a}{f_a})\left(\sigma_d\eta_u+\sigma_u\eta_d-\frac{\delta\pi}{2}\right)\right],
\end{gather}
\end{widetext}
where $\nu_k$ is the contribution of the condensate energy from the KMT interaction, and $\Sigma_k$ is the self-energy contributed by the KMT interaction.

\section{THERMODYNAMIC POTENTIAL}
\label{B}
According to finite-temperature field theory, the fermionic partition function is expressed via a path integral
\begin{equation}
    \mathcal{Z}=\int_{\text{anti}} \mathcal{D}[i\psi^\dagger]\mathcal{D}[\psi] \exp\left[\int_0^\beta d\tau \int d^3\vec{x}\, \mathcal{L}(\partial_0 \rightarrow i\partial_0)\right].
\end{equation}
Under the mean-field approximation, the Lagrangian density takes the form
\begin{equation}
\mathcal{L}_{\text{MF}} = \bar{\psi} \hat{S} ^{-1}\psi - \nu.
\end{equation}

Transforming into a momentum space with $p^\mu = (p_0, \vec{p})$ where $p_0 = (2n+1)i\pi T$, and replacing operators $\partial_\mu \rightarrow ip^\mu$, $\hat{S} ^{-1}\rightarrow S^{-1}(\vec{p},n)$, we derive the following
\begin{align}
\mathcal{Z} &= \int \mathcal{D}[\tilde{\psi}_n(\vec{p})] \mathcal{D}[i\tilde{\psi}^\dagger_n(\vec{p})] \exp\left\{ \beta \sum_{n,\vec{p}} \bar{\tilde{\psi}} S^{-1}(n,\vec{p}) \tilde{\psi} - \beta V\nu \right\} \notag \\
&= \prod_{\vec{p},n} \int d[\tilde{\psi}] d[i\tilde{\psi}^\dagger] \exp\left\{ i\tilde{\psi}^\dagger \gamma^0[-i\beta S^{-1}] \tilde{\psi} - \beta V\nu \right\} \notag \\
&= \exp(-\beta V\nu) \prod_{\vec{p},n} \det\left( \gamma^0[-i\beta S^{-1}] \right) \notag \\
&= \exp(-\beta V\nu) \prod_{\vec{p},n} \det(S^{-1}).
\end{align}
Using $\sum_{n,\vec{p}} = V\sum_n \int \frac{d^3\vec{p}}{(2\pi)^3}$, the partition function simplifies to
\begin{gather}
\ln\mathcal{Z} = -\beta V\nu + V\sum_n \int \frac{d^3\vec{p}}{(2\pi)^3} \ln\det(S^{-1}), \\
\Omega = \nu - T\int\frac{d^3\vec{p}}{(2\pi)^3}\sum_n \ln\det(S^{-1}).
\end{gather}

The determinant calculation of the matrix $S^{-1}$ proves computationally intensive. While Ref.~\cite{BoerBoomsma2008} employs numerical methods, we present analytical expressions revealing crucial mathematical structures beneficial for quark propagator calculations, similar to the approach in Refs.~\cite{Liu:2021gsi,Liu:2023uxm}.
Exploiting four-momentum summation symmetry $\sum_{n,\vec{p}}f(n,\vec{p}) = \sum_{n,\vec{p}}f(\pm n, \pm\vec{p})$
\begin{align}
\sum_{n,\vec{p}}\ln\det(S^{-1}(p)) &= \frac{1}{2}\sum_{n,\vec{p}}\left[\ln\det(S_1^{-1}(p_1)) + \ln\det(S_2^{-1}(p_2))\right] \notag \\
&= \frac{1}{2}\sum_{n,\vec{p}}\ln\det(S_1^{-1}(p_1)S_2^{-1}(p_2)).
\end{align}
Here, $p_2 = (\pm p_0, \pm\vec{p}_1)$ remains undetermined. We construct $S_2^{-1}$ to enhance symmetry while preserving the determinants
\begin{align}
S_2^{-1} &= \gamma^5\gamma^0 S_1^{-1}\gamma^0\gamma^5 \notag \\
= &\begin{pmatrix}
\slashed{p}_2 - \mu_u\gamma^0 - M_u - \beta_{ud}i\gamma^5 & S - Ri\gamma^5 \\
S - Ri\gamma^5 & \slashed{p}_2 - \mu_d\gamma^0 - M_d - \beta_{du}i\gamma^5
\end{pmatrix}.
\end{align}

Choosing $p_1 = -p_2$, we express
\begin{equation}
S_1^{-1}(p)S_2^{-1}(-p) = \begin{pmatrix}
A_{uu} & A_{ud} \\
A_{du} & A_{dd}
\end{pmatrix},
\label{B8}
\end{equation}
where
\begin{align}
  A_{uu} &= E_u^2 - (p_0 + \mu_u)^2 + \beta_{ud}^2 + S^2 + R^2, \\
  A_{ud} &= \begin{aligned}[t]
    & R(\beta_{ud} + \beta_{du}) - (M_u + M_d)S \\
    & + (S + Ri\gamma^5)(\mu_u - \mu_d)\gamma^0 \\
    & + \left[ (\beta_{ud} - \beta_{du})S + (M_u - M_d)R \right] i\gamma^5,
  \end{aligned} \label{eq:A_ud} \\
  A_{dd} &= A_{uu}(u \leftrightarrow d), \quad A_{du} = A_{ud}(u \leftrightarrow d). \label{eq:cross_terms}
\end{align}
The key properties $A_{uu} \propto I$ and $A_{ud}A_{du} \propto I$ yield the following
\begin{align}
  \ln\det(S^{-1}) 
  &= 2N_c \ln\Biggl\{ 
     \Bigl[\bigl(E_u^2 - (p_0 + \mu_u)^2 + \beta_{ud}^2 + S^2 + R^2\bigr) \cdot \notag \\
   &\bigl(E_d^2 - (p_0 + \mu_d)^2 + \beta_{du}^2 + S^2 + R^2\bigr)\Bigr] \notag \\
   &- S^2\Bigl[(M_u + M_d)^2 + (\beta_{ud} - \beta_{du})^2\Bigr] \notag \\
  &- R^2\Bigl[(M_u - M_d)^2 + (\beta_{ud} + \beta_{du})^2\Bigr] \notag \\
  + 4(\beta_{ud}&M_d + \beta_{du}M_u)SR + (S^2 + R^2)(\mu_u - \mu_d)^2 \Biggr\} \notag\\
  &= 2N_c \ln\biggl( a\left(p_0 + \frac{\mu_B}{3}\right)^4 + b\left(p_0 + \frac{\mu_B}{3}\right)^3 \notag \\
  & \quad+ c\left(p_0 + \frac{\mu_B}{3}\right)^2 + d\left(p_0 + \frac{\mu_B}{3}\right) + e \biggr) \notag\\
  &= 2N_c\sum_k \ln\left(p_0 + \frac{\mu_B}{3} - \lambda_k\right).
\end{align}
Here, $\lambda_k$ are roots of the quartic equation for $p_0 + \frac{\mu_B}{3}$, with coefficients $a$-$e$ derived from the expansion
\begin{align*}
    a&=1,\\
    b&=0,\\
    c&=-E_u^2-E_d^2-2R^2-2S^2-\beta_{du}^2-\beta_{ud}^2-\frac{\mu_I^2}{2},\\
    d&=\mu_I(M_u^2-M_d^2+\beta_{ud}^2-\beta_{du}^2),\\
    e&=(E_u^2+\beta_{ud}^2+S^2+R^2)(E_d^2+\beta_{du}^2+S^2+R^2)\\
   & +(S^2+R^2)(\mu_I)^2 -S^2\left[\left(M_u+M_d\right)^2+\left(\beta_{ud}-\beta_{du}\right)^2\right]\\
    &-R^2\left[\left(M_u-M_d\right)^2+\left(\beta_{ud}+\beta_{du}\right)^2\right]\\
   & +4\left(\beta_{ud}M_d+\beta_{du}M_u\right)SR.
\end{align*}
We can get the analytical expressions of the four roots $\lambda_k$
\begin{align*}
\lambda_1 &= +\frac{\sqrt{X}}{2} + \frac{1}{2} \sqrt{Y + \frac{Z}{4\sqrt{X}}}, \\
\lambda_2 &= -\frac{\sqrt{X}}{2} - \frac{1}{2} \sqrt{Y - \frac{Z}{4\sqrt{X}}}, \\
\lambda_3 &= +\frac{\sqrt{X}}{2} - \frac{1}{2} \sqrt{Y + \frac{Z}{4\sqrt{X}}},\\
\lambda_4 &= -\frac{\sqrt{X}}{2} + \frac{1}{2} \sqrt{Y - \frac{Z}{4\sqrt{X}}}. 
\end{align*}
with
\begin{gather*}
\Xi_1 = c^2 + 12e, \quad 
\Xi_2 = 2c^3 + 27d^2 - 72ce, \\
\Xi_3 = \left( \Xi_2 + \sqrt{-4\Xi_1^3 + \Xi_2^2} \right)^{\frac{1}{3}},\quad
\Xi = \frac{c}{3} + \frac{2^{\frac{1}{3}} \Xi_1}{3 \Xi_3} + \frac{\Xi_3}{3 \times 2^{\frac{1}{3}}}, \\
X = -c + \Xi, \quad 
Y = -c - \Xi, \quad 
Z = -8d.
\end{gather*}
Using the summation formula of the Matsubara frequencies,
\begin{equation}
    \sum_{n=-\infty}^{+\infty} \ln(p_0 \pm E)=\frac{\beta E}{2}+\ln(1+e^{-\beta E}).
\end{equation}
we can get the expression of the thermodynamic potential as Eq.~(\ref{thermo})

\section{QUARK CONDENSATE AND QUARK DENSITY}
\label{C}
In this appendix, we elegantly derive the quark propagator using the constructed $S_2^{-1}$.  
The inverse quark propagator matrix is expressed as
\begin{equation}
    S_1^{-1} = \begin{pmatrix}
        S_{0u}^{-1} & \Delta \\
        \Delta & S_{0d}^{-1}
    \end{pmatrix}.
\end{equation}
Using the critical properties $A_{uu} \propto I$ and $A_{ud}A_{du} \propto I$ from Eq.~(\ref{B8}), we construct
\begin{equation}
    S_2^{-1} = \begin{pmatrix}
        C_u S_{0u} & \Delta^\dagger \\
        \Delta^\dagger & C_d S_{0d}
    \end{pmatrix},
\end{equation}
with
\begin{gather}
    C_u = E_u^2 - (p^0 + \mu_u)^2 + \beta_{ud}^2, \\
    C_u S_{0u} = -\slashed{p} - \mu_u \gamma^0 - M_u - \beta_{ud} i\gamma^5, \\
    C_d = C_u(u \leftrightarrow d), \quad C_d S_{0d} = C_u S_{0u}(u \leftrightarrow d), \\
    \Delta^\dagger \Delta = S^2 + R^2.
\end{gather}

Recalling the determinant calculation process
\begin{align}
    \prod_k \Big(p^0& + \frac{\mu_B}{3} - \lambda_k\Big) = \det\left\{\begin{pmatrix}
        S_{0u}^{-1} &\Delta \\
        \Delta & S_{0d}^{-1}
    \end{pmatrix}
    \begin{pmatrix}
        C_u S_{0u} & \Delta^\dagger \\
        \Delta^\dagger & C_d S_{0d}
    \end{pmatrix}\right\} \notag \\
    =& \det\begin{pmatrix}
        C_u + \Delta^\dagger \Delta & S_{0u}^{-1}\Delta^\dagger + C_d \Delta S_{0d} \\
        S_{0d}^{-1}\Delta^\dagger + C_u \Delta S_{0u} & C_d + \Delta^\dagger \Delta
    \end{pmatrix} \notag \\
    = &C_u C_d + (\Delta^\dagger \Delta)^2 - \left(S_{0u}^{-1}\Delta^\dagger S_{0d}^{-1}\Delta^\dagger + \Delta C_d S_{0d} \Delta C_u S_{0u}\right) \notag \\
    = &\Big(S_{0u}^{-1} - \Delta S_{0d} \Delta\Big)\left(C_u C_d S_{0u} - \Delta^\dagger S_{0d}^{-1} \Delta^\dagger\right).
    \label{C7}
\end{align}
The structure $\Big(S_{0u}^{-1}\Delta^\dagger S_{0d}^{-1}\Delta^\dagger + \Delta C_d S_{0d} \Delta C_u S_{0u}\Big) $ $\propto I$ reflects intrinsic symmetry of the quark propagator
\begin{align}
    &\left(S_{0u}^{-1}\Delta^\dagger S_{0d}^{-1}\Delta^\dagger + \Delta C_d S_{0d} \Delta C_u S_{0u}\right) \notag \\
    =& \frac{1}{\Delta^\dagger \Delta} \Delta^\dagger \left(S_{0u}^{-1}\Delta^\dagger S_{0d}^{-1}\Delta^\dagger + \Delta C_d S_{0d} \Delta C_u S_{0u}\right) \Delta \notag \\
    = &\left(S_{0d}^{-1}\Delta^\dagger S_{0u}^{-1}\Delta^\dagger + \Delta C_u S_{0u} \Delta C_d S_{0d}\right).
\end{align}
Substituting $\left(S_{0u}^{-1}\Delta^\dagger S_{0d}^{-1}\Delta^\dagger + \Delta C_d S_{0d} \Delta C_u S_{0u}\right)$ into Eq.~(\ref{C7}), we obtain
\begin{equation}
    \prod_k \left(p^0 + \frac{\mu_B}{3} - \lambda_k\right) = \left(S_{0d}^{-1} - \Delta S_{0u} \Delta\right)\left(C_u C_d S_{0d} - \Delta^\dagger S_{0u}^{-1} \Delta^\dagger\right).
\end{equation}
Thus, the inverse propagator components are
\begin{equation}
    \left(S_{0u}^{-1} - \Delta S_{0d} \Delta\right)^{-1} = \frac{\left(C_u C_d S_{0u} - \Delta^\dagger S_{0d}^{-1} \Delta^\dagger\right)}{\prod_k \left(p^0 + \frac{\mu_B}{3} - \lambda_k\right)}.
\end{equation}

Using block matrix inversion
\begin{widetext}
\begin{equation}
    S = \begin{pmatrix}
        S_{uu} & S_{ud} \\
        S_{du} & S_{dd}
    \end{pmatrix} = \begin{pmatrix}
        \left(S_{0u}^{-1} - \Delta S_{0d} \Delta\right)^{-1} & -\left(S_{0u}^{-1} - \Delta S_{0d} \Delta\right)^{-1} \Delta S_{0d} \\
        -\left(S_{0d}^{-1} - \Delta S_{0u} \Delta\right)^{-1} \Delta S_{0u} & \left(S_{0d}^{-1} - \Delta S_{0u} \Delta\right)^{-1}
    \end{pmatrix},
\end{equation}
\end{widetext}
we derive
\begin{eqnarray}
    S_{uu} &=& \frac{\left(C_u C_d S_{0u} - \Delta^\dagger S_{0d}^{-1} \Delta^\dagger\right)}{\prod_k \left(p^0 + \frac{\mu_B}{3} - \lambda_k\right)}, \label{c12}\\
    S_{ud} &=& -\frac{\left(C_u C_d S_{0u} \Delta S_{0d} - \Delta^\dagger \Delta^\dagger \Delta\right)}{\prod_k \left(p^0 + \frac{\mu_B}{3} - \lambda_k\right)}, \\
    S_{dd} &=& S_{uu}(u \leftrightarrow d), \quad S_{du} = S_{ud}(u \leftrightarrow d).\label{c14}
\end{eqnarray}


Under trace operations, we utilize
\begin{equation}
    \frac{A(p_0)}{\prod_k \left(p^0 - \tilde{E}_k\right)} = \sum_{i=1}^4 \frac{1}{p^0 - \tilde{E}_i} \frac{A(\tilde{E}_i)}{\prod_{k \neq i} \left(\tilde{E}_i - \tilde{E}_k\right)} ,
\end{equation}
where $A(p_0)$ is a polynomial matrix in $p_0$ with $\text{Tr} A(p_0)$ being a polynomial of degree $<$ 4. 
We then have
\begin{equation}
    S_{q_1 q_2} = \sum_{i=1}^4 \frac{1}{p^0 - \tilde{E}_i} g_{q_1 q_2}(\tilde{E}_i)
\end{equation}
by defining
\begin{gather}
    g_{uu}(\tilde{E}_i) = \frac{\left(C_u C_d S_{0u} - \Delta^\dagger S_{0d}^{-1} \Delta^\dagger\right)\Big|_{p^0 = \tilde{E}_i}}{\prod_{k \neq i} \left(\tilde{E}_i - \tilde{E}_k\right)}, \\
    g_{ud}(\tilde{E}_i) = -\frac{\left(C_u C_d S_{0u} \Delta S_{0d} - \Delta^\dagger \Delta^\dagger \Delta\right)\Big|_{p^0 = \tilde{E}_i}}{\prod_{k \neq i} \left(\tilde{E}_i - \tilde{E}_k\right)}, \\
    g_{dd}(\tilde{E}_i) = g_{uu}(\tilde{E}_i)(u \leftrightarrow d), \quad g_{du}(\tilde{E}_i) = g_{ud}(\tilde{E}_i)(u \leftrightarrow d).
\end{gather}

Using the relations
\begin{widetext}
\begin{gather}
    \sigma_q = N_c \int \frac{d^3\vec{p}}{(2\pi)^3} T \sum_n \text{Tr}(S_{qq}) = 4N_c \sum_{i=1}^4 \int \frac{d^3\vec{p}}{(2\pi)^3} g_{\sigma q}(\tilde{E}_i) \left(-\frac{1}{2} + f(\tilde{E}_i)\right), \\
    \eta_q = N_c \int \frac{d^3\vec{p}}{(2\pi)^3} T \sum_n \text{Tr}(i S_{qq} \gamma^5) = 4N_c \sum_{i=1}^4 \int \frac{d^3\vec{p}}{(2\pi)^3} g_{\eta q}(\tilde{E}_i) \left(-\frac{1}{2} + f(\tilde{E}_i)\right), \\
    \pi = N_c \int \frac{d^3\vec{p}}{(2\pi)^3} T \sum_n \text{Tr}(i S_{ud} \gamma^5 + i S_{du} \gamma^5) = 4N_c \sum_{i=1}^4 \int \frac{d^3\vec{p}}{(2\pi)^3} g_{\pi}(\tilde{E}_i) \left(-\frac{1}{2} + f(\tilde{E}_i)\right), \\
    \delta = N_c \int \frac{d^3\vec{p}}{(2\pi)^3} T \sum_n \text{Tr}(S_{ud} + S_{du}) = 4N_c \sum_{i=1}^4 \int \frac{d^3\vec{p}}{(2\pi)^3} g_{\delta}(\tilde{E}_i) \left(-\frac{1}{2} + f(\tilde{E}_i)\right), \\
    \rho_q = N_c \int \frac{d^3\vec{p}}{(2\pi)^3} T \sum_n \text{Tr}(S_{qq} \gamma^0) = 4N_c \sum_{i=1}^4 \int \frac{d^3\vec{p}}{(2\pi)^3} g_{\rho q}(\tilde{E}_i) \left(-\frac{1}{2} + f(\tilde{E}_i)\right),
\end{gather}
\end{widetext}
where

\begin{eqnarray}
    g_{\sigma q}(\tilde{E}_i) &=& \frac{1}{4} \text{Tr}\left[g_{qq}(\tilde{E}_i)\right], \\
    g_{\eta q}(\tilde{E}_i) &=& \frac{1}{4} \text{Tr}\left[i g_{qq}(\tilde{E}_i) \gamma^5\right], \\
    g_{\pi}(\tilde{E}_i) &=& \frac{1}{4} \text{Tr}\left(i g_{ud}(\tilde{E}_i) \gamma^5 + i g_{du}(\tilde{E}_i) \gamma^5\right),
\end{eqnarray}
\begin{eqnarray}
    g_{\delta}(\tilde{E}_i) &=& \frac{1}{4} \text{Tr}\left(g_{ud}(\tilde{E}_i) + g_{du}(\tilde{E}_i)\right), \\
    g_{\rho q}(\tilde{E}_i) &=& \frac{1}{4} \text{Tr}\left[g_{qq}(\tilde{E}_i) \gamma^0\right],
\end{eqnarray}



and incorporating the quark propagator expressions Eqs.~(\ref{c12})-(\ref{c14}), we derive analytical forms for condensates as given in Eqs.~(\ref{35})–(\ref{40}).  
It can be rigorously shown that $\text{Tr}\left(C_u C_d S_{0u} - \Delta^\dagger S_{0d}^{-1} \Delta^\dagger\right)$ and $\text{Tr}\left(C_u C_d S_{0u} \Delta S_{0d} - \Delta^\dagger \Delta^\dagger \Delta\right)$ yield polynomials in $p^0$ of degree $<$ 3, satisfying Eq.~(\ref{mathe}).

\bibliography{ref}
\end{document}